\documentclass[journal]{IEEEtran}

\usepackage[ruled, vlined, linesnumbered]{algorithm2e}
\ifCLASSINFOpdf
\rmfamily
\usepackage{graphicx,subfigure}
\usepackage{amsmath}
\usepackage{array}
\usepackage{multirow}
\usepackage{amssymb}
\usepackage{dblfloatfix}
\usepackage{makecell}
\usepackage{bm}
\usepackage{color}
\usepackage{hyperref}
\begin{document}
\title{Allocation of Computation-Intensive Graph Jobs over Vehicular Clouds in IoV}

\author{%Michael~Shell,~\IEEEmembership{Member,~IEEE,}
%        John~Doe,~\IEEEmembership{Fellow,~OSA,}
%        and~Jane~Doe,~\IEEEmembership{Life~Fellow,~IEEE}% <-this % stops a space
Minghui LiWang, Seyyedali Hosseinalipour, Zhibin Gao, Yuliang Tang, Lianfen
Huang, \\Huaiyu Dai, \textit{Fellow, IEEE }
\thanks{\emph{{Minghui LiWang (minghuilw@stu.xmu.edu.cn), Zhibin Gao (corresponding author, gaozhibin@xmu.edu.cn), Yuliang Tang (tyl@xmu.edu.cn) and Lianfen Huang (lfhuang@xmu.edu.cn) are with the school of Information Science and Engineering, Xiamen University, Fujian, China. Seyyedali Hosseinalipour (shossei3@ncsu.edu) and Huaiyu Dai (hdai@ncsu.edu) are with the department of Electrical and Computer Engineering, North Carolina State University, NC, USA.}}}% <-this % stops a space
\thanks{\emph{}}}

\markboth{}%
{Minghui L.W. \MakeLowercase{\textit{et al.}}: Allocation of Computation-Intensive Graph Jobs over Vehicular Clouds in IoV}

\maketitle

\begin{abstract}
Graph jobs represent a wide variety of computation-intensive tasks in which
computations are represented by graphs consisting of components (denoting
either data sources or data processing) and edges (corresponding to data
flows between the components). Recent years have witnessed dramatic growth in smart vehicles and
computation-intensive graph jobs, which pose new challenges to the provision of
efficient services related to the Internet of Vehicles. Fortunately, vehicular clouds formed by a collection of
vehicles, which allows jobs to be offloaded among vehicles, can substantially
alleviate heavy on-board workloads and enable on-demand provisioning of
computational resources. In this paper, we present
a novel framework for vehicular clouds that maps components of graph jobs to
service providers via opportunistic vehicle-to-vehicle communication. Then,
graph job allocation over vehicular clouds is formulated as a 
non-linear integer programming with respect to vehicles' contact duration
and available resources, aiming to minimize job completion time and data
exchange cost. The problem is addressed for two scenarios: low-traffic and
rush-hours. For the former, we determine the optimal solutions for the problem. In the latter case, given the intractable
computations for deriving feasible allocations, we propose a novel low complexity randomized graph job allocation mechanism by considering
hierarchical tree based subgraph isomorphism. We evaluate the performance of our proposed algorithms through extensive simulations.
\end{abstract}


% \fi
\IEEEpeerreviewmaketitle

\section{Introduction}

\IEEEPARstart{T}{he} past decade has witnessed great advances in information and
communication technologies (ICTs). The proliferation of mobile traffic, smart mobile devices, and numerous related new applications have promoted the rapid development and widespread popularity of wireless communication networks (WCNs), as well as the Internet of Things (IoT). Smart devices such as smartphones and tablets with network access have experienced swift growth in number and variety, among which, smart vehicles are considered as the next
frontier in the automotive revolution, in which the number of interconnected
vehicles is predicted to reach 250~million by 2020~\cite{1}. As a result, the Internet of Vehicles (IoV)~\cite{2} is experiencing the significant transition of merging with wireless networks to create a powerful Intelligent Transportation System (ITS), where autonomous and semi-autonomous
driving represents a future trend~\cite{3}, \cite{4} . This technology promises to
substantially improve driving safety through intelligent operations such as
obstacle detection, collision avoidance, and experience of on-board entertainment 
including online games and augmented reality (AR)~\cite{5}.
Furthermore, technological advances in smart on-board equipment, including
computing processors and various sensing devices (e.g., on-board cameras and
high-quality sensors), can host perception-related applications with
innovative, computation-intensive features such as simultaneous localization and mapping (SLAM), user
face/gesture recognition~\cite{6}, as well as abnormal traffic identification and warning; such applications will greatly
influence the way people work and live in the future. Many of such applications can be characterized by graph jobs, wherein a job is represented by a graph, the
vertices (components)\footnote{A vertex in the graph job can be considered as a component describing a sub-job. We use “component” instead of “vertex” in the rest of this paper so as to accurately represent the physical significance.} of which denote either data sources or data processing, with
edges corresponding to data flows between vertices~\cite{7}-\cite{9}.
However, constraints related to computational resources and the capabilities
of on-board equipment~\cite{7}, \cite{10} pose major challenges to IoV, such
that the inherent limitation of a single smart vehicle can hinder the
fulfillment of job requirements.

To cope with the extensive and ever-changing application demands of smart
vehicles, mobile computation offloading (MCO) technology constitutes a new
paradigm allowing jobs to be offloaded to cloud servers for execution by
integrating communication and computing technologies~\cite{11}. Compared with remote
cloud servers, deploying multi-access edge computing (MEC)~\cite{12} servers located near access
points such as road side units (RSUs) can provide services that alleviate
severe transmission delays and performance degradation; however, users may still
experience signal coverage limitations (e.g., applications such as emergency medical diagnosis in disaster scenes where infrastructures are destroyed and remote regions such as 
mountains and deserts without RSU coverage) and resource constraints, especially during high-traffic periods. As a result, the vehicular cloud (VC)~\cite{13} has
been introduced recently as a type of mobile device cloud (MDC), where vehicles with idle resources, named ``service
providers'' (SPs), within a job owner's (JO's) communication range can act
as mobile computing servers and form a cloud mainly by leveraging
vehicle-to-vehicle (V2V) communication technology~\cite{14}, \cite{15} to process
the job in parallel. This technology significantly relieves heavy on-board
workloads for JOs and accelerates job completion. The feedback duration of computation results is often ignored given that their data size is much smaller than that of the application data~\cite{10}, \cite{11}; alternatively, results can be transferred back to the JO via a multi-hop V2V routing
path or uploaded to RSUs for future
delivery in case the V2V connection is disrupted.

In this paper, we introduce a novel VC-based IoV framework. Without loss of generality, it is assumed that each VC
contains one JO with a graph job and a collection of SPs that can interact
with the JO via one-hop V2V communications. Moreover, the V2V communication
configuration of each VC supports data flows corresponding to graph edges.
We utilize slot-based\footnote{An idle slot is seen as a resource block that can run one component of a graph job~\cite{8}, \cite{9}.} 
representation to quantify the available
resources at each SP~\cite{9}, where the number of available slots
differs from one vehicle to another based on their heterogeneous resource
capabilities. The job completion time and data exchange cost represent key concerns in designing the principal mechanism for
allocation of computation-intensive graph jobs over VCs, the latter of which
occurs when two connected job components are mapped to slots of different
SPs. In our proposed allocation mechanism, each job component is efficiently mapped
(offloaded) to an applicable idle slot of a SP under constraints of
opportunistic contacts and available resources. 

Our main contributions can be summarized as follows:
\begin{enumerate} 
\item We first establish an IoV framework containing VCs, which enables
computation-intensive graph jobs to be mapped (offloaded) to SPs to overcome
limitations in the on-board resources and computational capabilities
of JOs by capitalizing on the opportunistic contact duration between
vehicles.

\item We formulate the graph job allocation
problem as a nonlinear integer programming (NIP) problem under constraints
of opportunistic contacts between SPs and available slots
aiming to minimize the job completion
time and data exchange cost.

\item To tackle the aforementioned NIP problem, we first focus on low-traffic
IoV scenarios, for which we develop a graph job allocation algorithm
to find the optimal solution. This approach relies on
addressing the sub-graph isomorphism problem\footnote{The subgraph isomorphism problem is a computational task in which two graphs H1 and H2 are given as the input, and one must determine whether H1 contains a subgraph that is isomorphic to H2.}, which is known to be NP-complete~\cite{8,22}.
This makes our first proposed algorithm ineffective upon having a high vehicular density or equivalently large network size. To address this issue, we propose a
 randomized graph job allocation mechanism by considering hierarchical tree based subgraph isomorphism, which enjoys a low computational complexity, and conforms well to fast-changing environments in IoV.
 
\item We conduct a thorough numerical analysis to evaluate the performance of
the proposed algorithms under different graph job topologies, traffic
densities, vehicular cloud configurations and numbers of idle slots.
\end{enumerate}

The rest of this paper is organized as follows: after describing the
motivation, feasibility, and related work in Section II, we introduce system models in Section III. The problem formulation is given in
Section IV. Then, in Section V and VI, the optimal allocation mechanism and the hierarchical tree based randomized allocation mechanism are presented, respectively. 
We evaluate the performance of the proposed mechanisms  in Section VII, and provide conclusion, challenges, as well as future works in Section VIII.

\section{Background and Related work}

\subsection{Background}
\noindent  The increasing demands on autonomous driving coupled
with the vast amount of data collected by rapidly developed on-board
equipment has bolstered the popularity of computation-intensive
applications. However, inherent drawbacks (such as limited computational
resources and capability, as well as perceptive sensor range) of a single
smart vehicle can often result in unsatisfactory computational performance
of such applications~\cite{16}. Moreover, constraints on wireless coverage of base
stations (BSs) and RSUs, especially in remote regions, render it difficult
for smart vehicles to readily enjoy services from cloud computing servers.

VCs can efficiently realize coordination among smart
vehicles mainly by utilizing V2V communications, taking full advantage of
opportunistic connections, and exploiting idle resources in fast-changing
topologies~\cite{13,14}.

\begin{figure*}[!t]\centering	
    \subfigure[]{\includegraphics[width=3in]{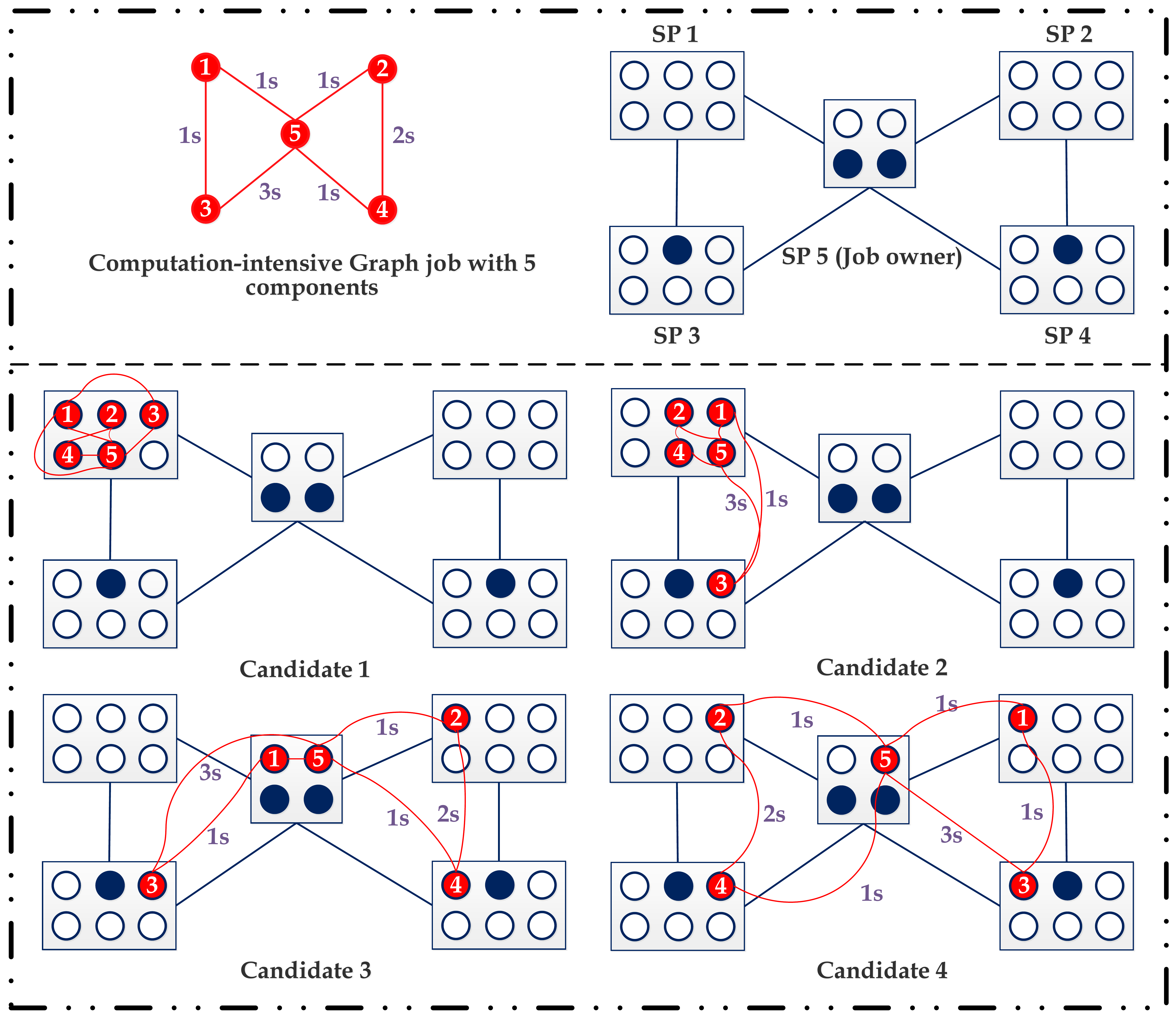}}\label{fig1(a)}\qquad\qquad\qquad
    \subfigure[]{\includegraphics[width=3in]{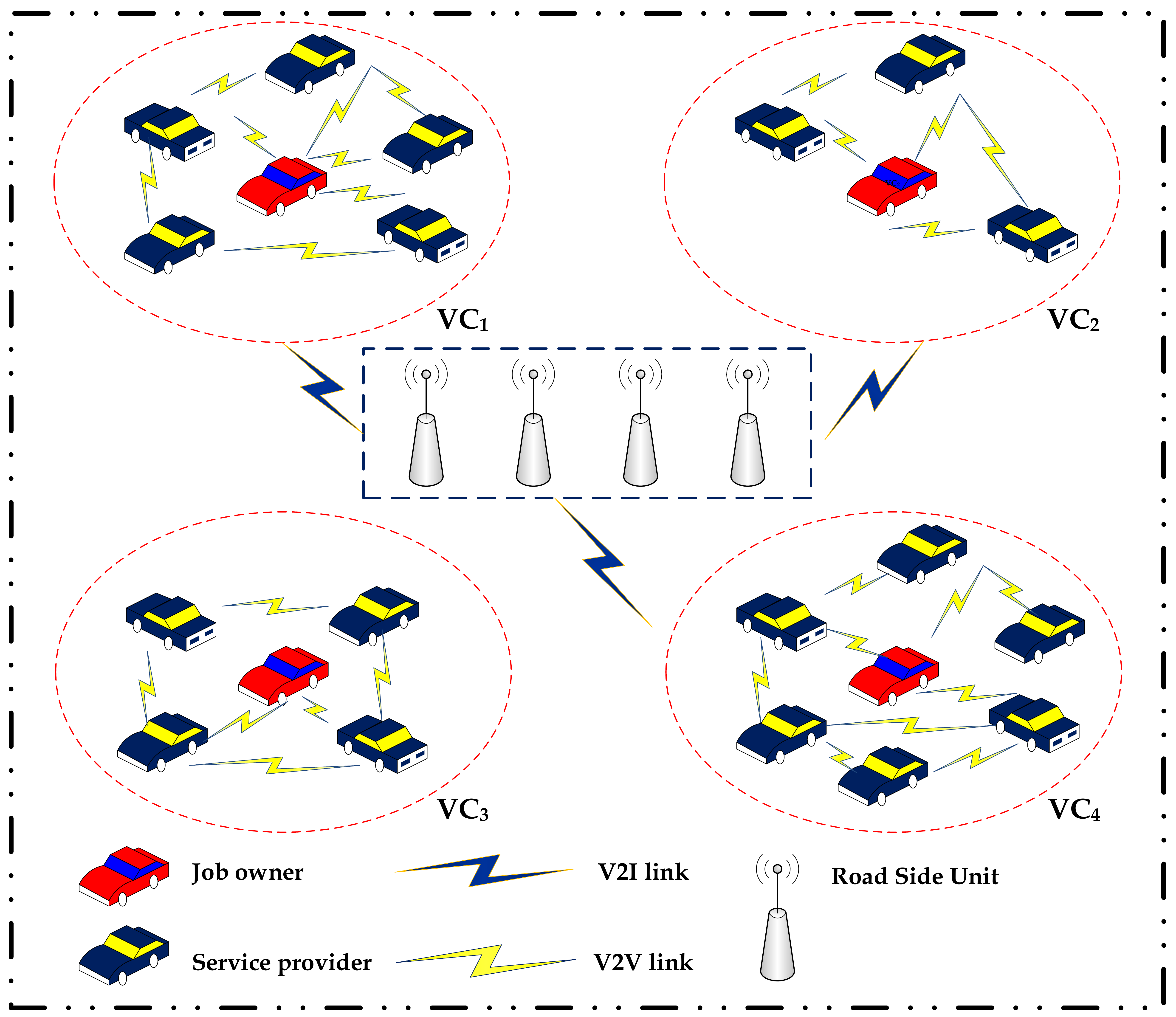}}\label{fig1(b)}
\caption{(a) illustrative candidate examples
for partitioning and distributing a 5-component graph job in a VC of one JO and four SPs with different numbers of slots (solid circles and hollow circles denote
occupied slots and idle slots, respectively). In this stylized example, all
edges that indicate two connected components scheduled on different SPs
incur data exchange costs. The costs of partitioning the job according to
these candidate examples are ${Cost}_{\mathrm{candidate}1}=0, {
Cost}_{\mathrm{candidate}2}=2c_{13}^{Exch},
{Cost}_{\mathrm{candidate}3}=2c_{35}^{Exch}+c_{25}^{Exch}+c_{45}^{Exch}+c_{24}^{Exch},
{Cost}_{\mathrm{candidate}4}=c_{13}^{Exch}+c_{15}^{Exch}+c_{25}^{Exch}+c_{35}^{Exch}+c_{45}^{Exch}+c_{24}^{Exch}$; (b) Vehicular clouds scenario. }
\label{fig1}
\end{figure*}

\subsection{Related work}
\noindent The increasing popularity of computation-intensive applications offers great convenience; however, such applications usually place a substantial burden on on-board equipment in IoV, often going beyond the capability of a smart vehicle. This fact has resulted in a growing interest in studying efficient computation offloading strategies. The existing literatures on computation-intensive application offloading can be roughly divided into two categories: 1) those where applications are directly mapped as bit streams and considered as a collection of sub-applications without considering the their dependencies, such as \cite{17}-\cite{19}; 2) those explicitly considering the structure of applications which can be modeled as directed/undirected graphs, such as \cite{6},\cite{7},\cite{9},\cite{10},\cite{20},\cite{21}. A reliability-oriented stochastic optimization
model in vehicle-infrastructure systems is proposed in~\cite{17} based on the dynamic programming for computation offloading considering the deadline constraint on application execution. A hierarchical IoV framework with backup computing servers is introduced in~\cite{18}, where a Stackelberg game theoretic approach is adopted to design an optimal multilevel offloading scheme, which maximizes the utilities of both the vehicles and the computing servers. The multiple-vehicle computation offloading in vehicular edge networks is studied in~\cite{19} and formulated as a multi-user computation offloading game, where the existence of Nash equilibrium (NE) is proved and a distributed algorithm is proposed to compute the equilibrium. In the aforementioned works, the inner structure of the applications is ignored, which has become more and more important when designing computation offloading strategies.

Over the past years, only a few studies have
investigated the allocation of graph-based computation-intensive applications. In the context of cloud computing assuming a static network topology, a low-complexity randomized scheduling algorithm without service preemptions is studied in~\cite{9}, which stabilizes the system with graph arrivals/departures, and thus facilitates a smooth tradeoff between minimizing
the average partitioning cost and average queue length.
In the context of mobile computing environments, where service providers are static, a dynamic offloading algorithm for
directed-graph-based jobs based on Lyapunov optimization is presented in~\cite{6}, which satisfies the
requirements of energy conservation and application execution time. Applications are modeled by directed taskgraphs in~\cite{7} where nodes represent tasks and edges indicate the dependency of tasks. Both sequential and concurrent task offloading algorithms are proposed to minimize the application completion time. It is assumed in~\cite{10} that each mobile device has several independent tasks as a set and the task offloading scheduling and transmit power allocation in MEC systems are jointly optimized. Moreover, a low-complexity sub-optimal algorithm is proposed to minimize the weighted
sum of the execution delay and device energy consumption based on
alternating minimization. The authors in~\cite{20} focus on scheduling
parallel jobs composed of a set of independent tasks and
consider energy consumption while minimizing job completion time.
A fast hybrid multi-site computation offloading mechanism is proposed in~\cite{21}, which finds the offloading solution in a timely manner by considering the application size, where applications are modeled as weighted relation graphs. 

The rapid growth of the automobile~industry has brought about more and more on-board
resources in vehicles, which can be dynamically reallocated to support opportunistic job allocations. However, the aforementioned works paid inadequate attention to on-board resource utilization and optimization and users' mobilities especially for service providers, which is highly relevant in IoV, where high vehicular mobility leads to difficulties in opportunistic contact
between vehicles in fast-changing network topologies. To the best of our knowledge,  this paper is among the first to study the allocation of computation-intensive graph
jobs over VCs while considering their inherent characteristics, where the graph job allocation problem over vehicular clouds is abstracted as a novel subgraph isomorphism problem under opportunistic communication constraints.

\section{Problem Overview and System Model}
\subsection{Problem Overview}

We propose a framework in which graph jobs can be offloaded to several SPs in the relevant VC via
one-hop V2V channels. Different SPs have different numbers of idle slots,
and each slot can process one component of a graph job. Note that structural
characteristics exist in the offloaded computational jobs and the IoV topology; thus, we model both offloaded jobs
and VCs as weighted undirected graphs. Under constraints of limited
opportunistic contact duration between vehicles and idle resources, each VC
aims to effectively allocate all the components of a graph job to available SPs while
minimizing job completion time and data exchange cost.

\subsection{Vehicular Contact Model}

An opportunistic contact event between vehicle $j$ and $j'$ occurs during
$\tau\mathrm{\in (}\tau_{\mathrm{1}},\tau_{\mathrm{2}}\mathrm{)}$ if the
following conditions are satisfied: $ \|L_{j}( \tau_{\mathrm{1}})-L_{j'}\mathrm{(}\tau_{\mathrm{1}}\mathrm{)} \|>R$,  $\|L_{j}( \tau )-L_{j'}\mathrm{(}\tau \mathrm{)}\|\leq R$ and $\| L_{j}( \tau_{\mathrm{2}})-L_{j'}\mathrm{(}\tau_{\mathrm{2}}\mathrm{)}\|>R$, \noindent where $L_{j}\mathrm{(}\tau \mathrm{)}$ and
$L_{j'}\mathrm{(}\tau \mathrm{)}$ denote vehicular locations at time $\tau$, $\|\cdot\|$ represents the Euclidean distance, and $R$
indicates the communication radius of vehicles. It is assumed that the contact
duration between two vehicles obeys the exponential distribution~\cite{23} with
parameter ${\mathrm{ \lambda }}_{jj'}$; hence, the probability that the
contact duration $\mathrm{\Delta }\tau_{jj'}$ between vehicle $j\mathrm{
}$ and $j'$ is larger than $T$, $T>0$ can be calculated as
$P( \Delta \tau_{jj'}>T)=e^{-{T\lambda }_{jj'}}$. Correspondingly, the larger the value of this probability is, the more ensurance can be achieved to protect successful data transmission between moving vehicles.

\subsection{Model of Graph Jobs}

Consider a collection of graph jobs ${\bm J}$ with different types, each of which can be described as a graph\textbf{
}${\bm G}_{{\bm a}}^{{\bm A}}=({\bm V}_{{\bm a}}^{{\bm A}},{\bm E}_{{\bm a}}^{{\bm A}},{\bm W}_{{\bm a}}^{{\bm A}})$, containing a set of components (computations)
${\bm V}_{{\bm a}}^{{\bm A}}=\{\upsilon
_{i}^{A_{a}}|i\in {\bm N}_{{\bm a}}\}$,\linebreak\\[-11pt] where each component requires a
slot, along with a set of edges\textbf{
}${\bm E}_{{\bm a}}^{{\bm A}}=\{e_{ii'}^{A_{a}}|i\in
{\bm N}_{{\bm a}}, \mathbf{ }i'\in {\bm N}_{{\bm a}}\mathbf{
}, i\ne i'\}$ with associated weight\textbf{
}${\bm W}_{{\bm a}}^{{\bm A}}=\{\omega_{ii'}^{A_{a}}|i\in
{\bm N}_{{\bm a}}, i'\in{\bm N}_{{\bm a}}\mathbf{
}, i\ne i'\}$, and $\left| {\bm N}_{{\bm a}} \right|$ denotes
the number of components of job type $a \in  \bm {J}$. Each graph
${\bm G}_{{\bm a}}^{{\bm A}}$ represents how the
computation is supposed to be splitted among components in
${\bm V}_{{\bm a}}^{{\bm A}}$. In the above notations, the superscript $\bm A$ denotes the application. Edges represent data flows
between these components; the weight $\omega_{ii'}^{A_{a}}$ of edge
$e_{ii'}^{A_{a}}$ indicates the requested connecting duration between
components $\upsilon_{i}^{A_{a}}$ and $\upsilon_{i'}^{A_{a}}$. In other words, the contact duration of SPs that are handling components $\upsilon_{i}^{A_{a}}$ and $\upsilon_{i'}^{A_{a}}$ should be equal to or larger than $\omega_{ii'}^{A_{a}}$.
An example of a graph job with five components is illustrated in Fig. 1(a).

\subsection{Model of Vehicular Clouds}

Consider a scenario with $|{\bm O}|+|{\bm S}|$ vehicles moving on
north-south and east-west roads, where $o_{i}\in
{\bm O}$ are JOs and $s_{j}\in
{\bm S}$ are SPs, each of which has a different number of idle
slots. In one snapshot, vehicles are partitioned into
$|{\bm O}|$ VCs, ${\bm Q} = \{ {VC}_{1}, {VC}_{2},...,{VC}_{|{\bm O}|} \}$. Each ${VC}_{i}\in {\bm Q}$ contains one JO $o_{i}$
and several SPs, all of which can communicate with $o_{i}$ via a one-hop V2V
channel. Notably, because $o_{i}$ may have idle slots, each JO is
considered as a virtual SP; as such, each VC can be regarded as an undirected
connected graph. In this article, ${VC}_{i}$ is considered to
provide computational service to process the graph job owned by
$o_{i}$ in parallel. Suppose that there is no interference among VCs, 
we take one VC in ${\bm Q}$ as an example in analyzing graph job allocation mechanisms in different traffic scenarios.

Consider a VC containing a collection of SPs ${\bm M}$, where each SP $s_{j}\in {\bm M}$\textbf{
}has a set of idle slots $\mathbf{\bm \kappa }_{{j}}$\textbf{
}that are fully connected with each other and can be used to run
up to $\mathrm{\mathbf{|}}{\bm \kappa
}_{{j}}\mathbf{|}$ processes (components) in parallel. For
notational simplicity, $m=|{\bm M}|$ represents the number of SPs (including JO) in this VC and $s_{m}$ denotes the virtual SP
corresponding to the JO. The VC can be represented as a graph
${\bm G}^{{\bm S}}=({\bm V}^{{\bm S}},{\bm E}^{{\bm S}},{\bm W}^{{\bm S}})$, consisting of a SP set ${\bm V}^{{\bm S}}=\{s_{j}|s_{j}\in
{\bm M}\}$; an edge set ${\bm E}^{{\bm S}}=\{e_{jj'}^{S}|s_{j}\in
{\bm M},s_{j'}\in {\bm M},j\ne j'\}$, where each $e_{jj'}^{S}$ takes a binary value indicating that
$s_{j}$ can communicate with $s_{j'}$ via a one-hop V2V
channel ($e_{jj'}^{S}=1$) or not ($e_{jj'}^{S}=0)$; and the corresponding weight set ${\bm W}^{{\bm S}}=\{\lambda
_{jj'}|s_{j}\in {\bm M},s_{j'}\in {\bm M},j\ne j'\}$ where $\lambda_{jj'}$ denotes the parameter of the corresponding exponential distribution on contact duration between $s_j$ and $s_{j'}$. The superscript $\bm S$ in the notations denotes the service. Please note that there are two graph models considered in this study, one for the structure of the graph jobs discussed in III.C and the other for the topology of VC discussed here. Each idle slot can run one component of a graph job and the execution time of which is denoted as $t^{Exec}$, which is assumed to be the same for all components of the graph job. The proposed model for VCs
is depicted in Fig. 1(b).

\subsection{Candidate}

An important construct in this article is the concept of a candidate. For
any job type, there are several ways (an exponentially large number) in
which a job can be distributed over the VC as depicted in Fig. 1(a). A
candidate corresponds to a sub-graph with the same structure as the job in a
VC, which satisfies the requested connecting durations for all pairs of connected components.

\subsection{Data Exchange Cost}

We assume the cost of data exchange between two slots on the same service
provider to be zero; otherwise, $c_{jj'}^{Exch}$ is defined as the cost if
two connected components are allocated to slots of different SPs, $s_{j}$
and $s_{j'}$. This captures the cost incurred from the total
traffic exchange among different SPs in one VC. Several examples of the data
exchange cost are presented in Fig. 1(a).

\section{Problem Formulation of Computation-Intensive Graph Job Allocation Over Vehicular Clouds}

\SetKwInOut{Input}{Input}\SetKwInOut{Output}{Output}

\noindent To better analyze the framework of graph job allocation, we mainly study the
problem in one VC since the proposed mechanism is universal across all job
types and VCs. Hence, we use simpler notations in which a graph job is represented as:
${\bm G}^{{\bm A}}=({\bm V}^{{\bm A}}$, ${\bm E}^{{\bm A}}$, ${\bm W}^{{\bm A}})$
that contains components set ${\bm V}^{{\bm A}}=\{\upsilon_{i}^{A}|i\in \textbf{\textit{N}}\}$; edge set ${\bm E}^{{\bm A}}=\{e_{ii'}^{A}|i\in
{\bm N}$, $i'\in {\bm N}, \mathbf{ }i\ne i'\}$ and the
associated weight set ${\bm W}^{{\bm A}}=\{\omega_{ii'}^{A}|i\in
{\bm N}, i'\in {\bm N},i\ne i'\}$, where $n=\left| {\bm N}
\right|$ denotes the total number of components in the graph job. Moreover, a VC is
defined as graph
${\bm G}^{{\bm S}}=({\bm V}^{{\bm S}},{\bm E}^{{\bm S}},{\bm W}^{{\bm S}})$
with SP set ${\bm V}^{{\bm S}}$, edge set
${\bm E}^{{\bm S}}$, and a weight set corresponding to edges
${\bm W}^{{\bm S}}$, as mentioned in Section III.D. We
denote the data transmission duration of one
component from $s_m$ (the JO) to $s_{j}$ as $t_{jm}^{Trans}$ ($1\leq j \le m$), which is related to channel conditions, packet loss and retransmissions as well as transmission
power. Consider the binary indicator $x_{ij}$ for which $x_{ij}=1$ if component $\upsilon_{i}^{A}$ is
assigned to $s_{j}$, and $x_{ij}=0$, otherwise. As for the
cost model of job partitioning among different SPs, let the binary indicator
$y_{jj'}=1$ denote the existence of data exchange between $s_{j}$ and
$s_{j'}$; otherwise, $y_{jj'}=0$.

Based on the above notations, the
indicator $y_{jj'}$ can be represented as a piecewise function of $x_{ij}$ and
$x_{i'j'}$ shown as equation (1).

\begin{equation}
y_{jj'}=\left\{
\begin{array}{rcl}
1, &  {\forall e_{ii'}^{A}\in {\bm E}^{{\bm A}}, if~j\ne j'~and~{x}_{ij}\cdot x_{i'j'}=1 },\\
0, &  {otherwise},\\
\end{array} \right.
\end{equation}

%\label{eq1}
%&\mathop{{\rm min}}\limits_{ {\bm X}}~U^{t}\left({\bm X} \right)+U^{c}({\bm Y(\bm X)})\\
%& \textit {s.t.} \nonumber \\
%$\forall e_{ii'}^{A}\in {\bm E}^{{\bm A}}$, if $j\ne j'$ and
%${x}_{ij}\cdot x_{i'j'}=1$;
%
%\quad
%\nonumber
%\end{align}

\noindent The job completion time can be calculated as:
%\begin{equation}
%U^{t}( {{ \bm X}} )={\rm max}(
%\sum\nolimits_{i=1}^n {x_{i1}t_{1m}^{Trans}},..,\sum\nolimits_{i=1}^n {x_{im-1}t_{m-1m}^{Trans}}
%)\nonumber  \\
%+t^{Exec}
%\end{equation}

\begin{equation}
U^{t}( {{ \bm X}} )={\rm max}(
[\sum\nolimits_{i=1}^n {x_{ij}t_{jm}^{Trans}} ]_{{1\leq  j\leq m-1}})+t^{Exec},
\end{equation}

\noindent where ${\bm X}=[ x_{ij} ]_{{1\leq i\leq
n,1\leq  j\leq m}}$ defines the matrix of indicator\textbf{
}$x_{ij}$, the first term describes the data transmission
duration given that each JO can offload different numbers of components to
different SPs in parallel and the second term represents the job execution
time, which is introduced in Section III.D. We can then formulate the total cost of job allocation as the following function::

\begin{equation}
U^{c}({\bm Y(\bm X)})=\frac{1}{2}\cdot \sum\nolimits_{j=1}^m
\sum\nolimits_{j'=1}^m {y_{jj'}\cdot } c_{jj'}^{Exch},
\end{equation}

\noindent  where ${\bm Y(\bm X)}\mathrm{\mathbf{=}}\left[ y_{jj'} \right]_{{1\leq j\leq
 m,1\leq }{ j}^{\mathbf{'}}{\bm \leq m}}$ denotes the
matrix of indicator ${ y}_{jj'}$, which is a function of $\bm X$ according to (1). 

\begin{figure*}[!t]\centering
	\subfigure[]{\includegraphics[width=2.35in]{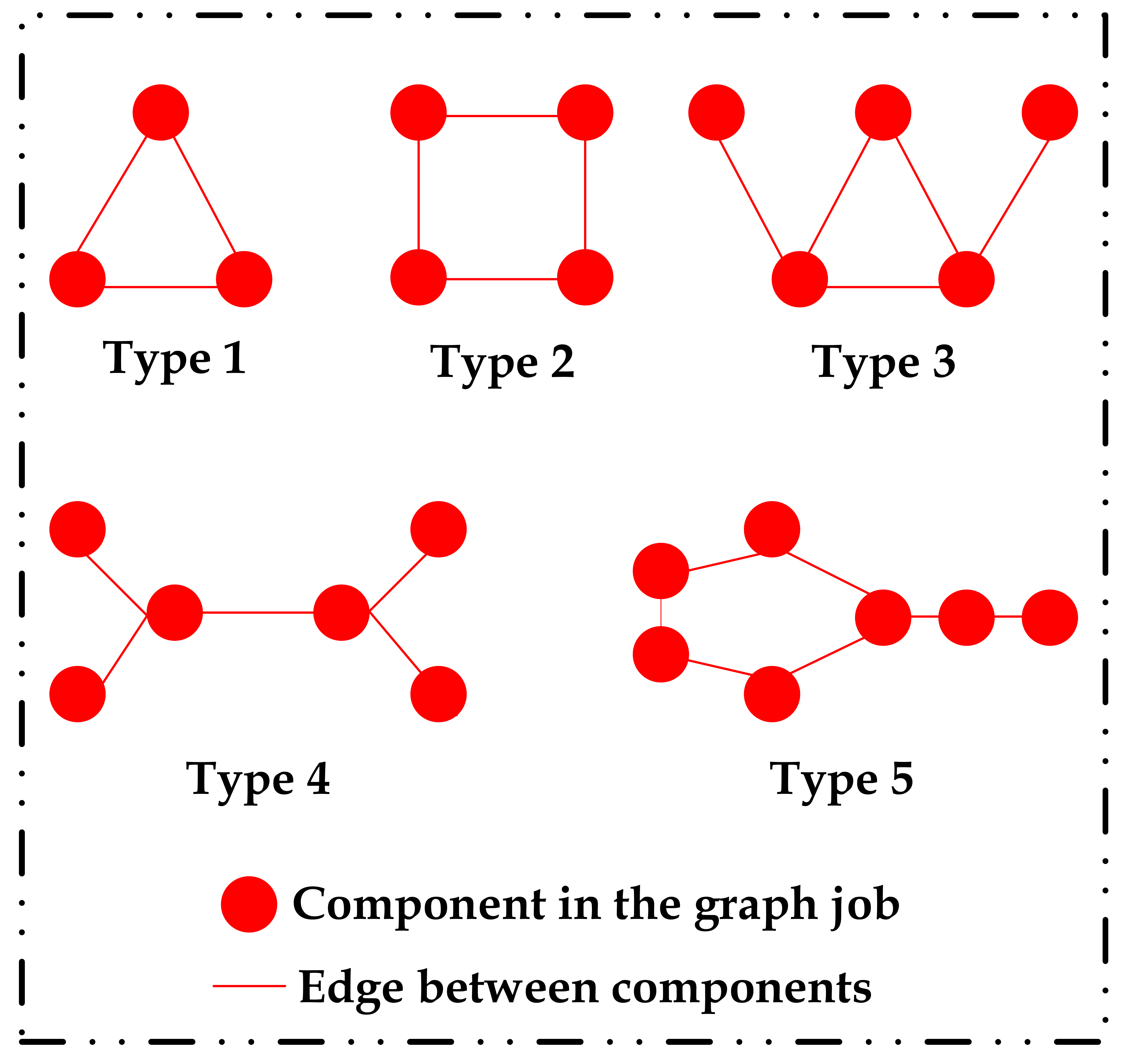}}\qquad\qquad\qquad
	\subfigure[]{\includegraphics[width=3.6in]{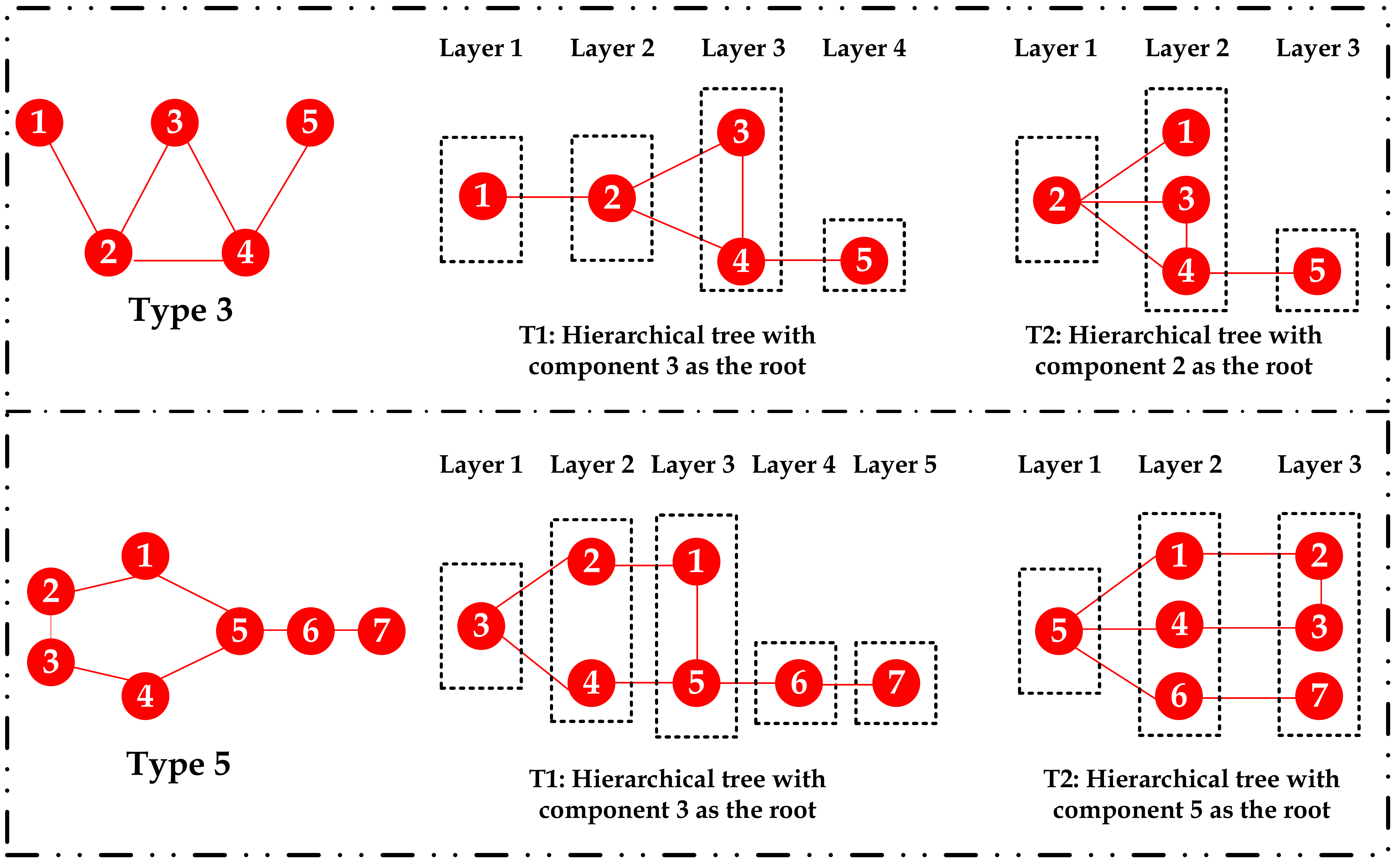}}
	\caption{Different graph job topologies and examples on hierarchical trees corresponding to different graph jobs: a) different graph job topologies: Type 1 - closed triad graph, Type 2 - square graph, Type 3 - bull graph, Type 4 - double-star graph, Type 5 - tadpole graph; b) examples on hierarchical trees with layers corresponding to job Type 3 and Type 5.}
	\label{fig2}
\end{figure*}

Correspondingly, we formulate the graph job allocation problem in a VC, aiming to minimize the sum of job completion time and data exchange cost under opportunistic contact and idle resource constraints, as follows:
%\begin{align}
%\label{eq4}
%&\mathop{{\rm min}}\limits_{ {\bm X}}~U^{t}\left({\bm X} \right)+U^{c}({\bm Y(\bm X)})\\
%
%&  {s.t.} \nonumber \\
%&(a).\quad \forall s_{j}\in {\bm M},~\sum\limits_{i=1}^n x_{ij} \leq
%\mathrm{\mathbf{|}}\mathbf{\bm \kappa }_{\bm j}\mathbf{|}\nonumber\\
%
%
%&(b).\quad \forall e_{ii'}^{A}\in {\bm E}^{{\bm A}}, \mbox{\textit{if}}~s_j\ne
%s_j'~\mbox{\textit{and}}~x_{ij}\cdot x_{i'j'}=1,\nonumber\\
%&\qquad {e}^{{-({|t}_{jm}^{Trans}-t_{j'm}^{Trans}|+\omega_{ii'}^{A})}\lambda
%_{jj'}}\geq \varepsilon\nonumber\\
%
%
%&(c).\quad \forall x_{ij}=1~\mbox{\textit{and}}~{s_j\ne s_m, e}^{-\lambda_{jm}\sum\limits_{i=1}^n
%{x_{ij}t_{jm}^{Trans}} }\geq \xi\nonumber\\
%
%
%&(d).
%\quad
%\sum\limits_{j=1}^m \sum\limits_{i=1}^n {x_{ij}=n}\nonumber
%\end{align}

%\begin{align}
%\label{eq4}
%&\mathop{{\rm min}}\limits_{ {\bm X}}~U^{t}\left({\bm X} \right)+U^{c}({\bm Y(\bm X)})\\
%& \textit {s.t.} \nonumber \\
%&(a).\quad \sum\limits_{i=1}^n x_{ij} \leq
%\mathrm{\mathbf{|}}\mathbf{\bm \kappa }_{\bm j}\mathbf{|}, \forall s_{j}\in {\bm M}  \nonumber\\
%%\end{align}
%%\begin{align*}
%&(b).\quad \forall e_{ii'}^{A}\in {\bm E}^{{\bm A}}, \mbox{\textit{if}}~s_j\ne
%s_j'~\mbox{\textit{and}}~x_{ij}\cdot x_{i'j'}=1,\nonumber\\
% {e}^{{-({|t}_{jm}^{Trans}-t_{j'm}^{Trans}|+\omega_{ii'}^{A})}\lambda
%_{jj'}}\geq \varepsilon\nonumber\\
%&(c).\quad {e}^{-\lambda_{jm}\sum\limits_{i=1}^n
%{x_{ij}t_{jm}^{Trans}} }\geq \xi, \forall x_{ij}=1~\mbox{\textit{and}}~{s_j\ne s_m} \nonumber\\
%&(d).
%\quad
%\sum\limits_{j=1}^m \sum\limits_{i=1}^n {x_{ij}=n}\nonumber
%\end{align}

\begin{align}
\label{eq4}
&\mathop{{\rm min}}\limits_{ {\bm X}}~\alpha_1 U^{t}\left({\bm X} \right)+\alpha_2 U^{c}({\bm Y(\bm X)})\\
& \textit {s.t.} \nonumber \\
&(a).\quad \sum\limits_{i=1}^n x_{ij} \leq
\mathrm{\mathbf{|}}\mathbf{\bm \kappa }_{\bm j}\mathbf{|}, \forall s_{j}\in {\bm M},  \nonumber\\
%\end{align}
%\begin{align*}
&(b).\quad {e}^{{-({|t}_{jm}^{Trans}-t_{j'm}^{Trans}|+\omega_{ii'}^{A})}\lambda_{jj'}}\geq \varepsilon, \nonumber\\ 
&\qquad~\forall e_{ii'}^{A}\in {\bm E}^{{\bm A}}, \mbox{\textit{if}}~s_j\ne s_j'~\mbox{\textit{and}}~x_{ij}\cdot x_{i'j'}=1,\nonumber\\
&(c).\quad {e}^{-\lambda_{jm}\sum\limits_{i=1}^n
{x_{ij}t_{jm}^{Trans}} }\geq \xi, \forall x_{ij}=1,~{j\ne m},\nonumber\\
&(d).
\quad
\sum\limits_{j=1}^m \sum\limits_{i=1}^n {x_{ij}=n},\nonumber
\end{align}

\noindent where $\alpha_1$ and $\alpha_2$ denote non-negative weight parameters which describe the preference between job completion time and data exchange cost, respectively, with $\alpha_1+\alpha_2=1$. In this formulation, constraint $(a)$ imposes restrictions on idle resources where $| \bm{\kappa}_{\bm j}|$ denotes the number of idle slots on each SP.
Since the contact duration between vehicles follows exponential distribution as mentioned in the aforementioned section, constraint $(b)$ and $(c)$ are probabilistic constraints, where $(b)$ ensures that if two connected components $\upsilon
_{i}^{A}$ and $\upsilon_{i'}^{A}$ with edge $e_{ii'}^{A}$ and associated weight $\omega_{ii'}^{A}$ are
allocated to different SPs $s_{j}$ and $s_{j'}$, respectively, the
probability of the contact duration between $s_{j}$ and $s_{j}$ being larger
than $\omega_{ii'}^{A}$ should be greater than threshold $ \varepsilon $. Notably, $|{t}_{jm}^{Trans}-t_{j'm}^{Trans}|+ \omega_{ii'}^{A}$ is the total time that adjacent vehicles need to maintain their contact, which consists of the effective transmission time and processing time.
Similarly, constraint $(c)$ confirms successful data transmission of each component from JO to SPs, where $\xi $ denotes a positive constant which is less than or equal to 1. Notably, the larger the values of $ \varepsilon $ and $\xi $ are, the higher guarantees can be met for the graph job allocation requirements.
Also, constraint $(d)$ ensures the assignment of all the components to available slots in the related VC. In this study, it is assumed that the system parameters such as $t_{j'm}^{Trans}$, $\lambda_{jj'}$ and $| \bm{\kappa}_{\bm j}|$ involved in (4) are average statistics and will stay unchanged during graph job allocation~\cite{23}\cite{24}.

\begin{algorithm}
\label{algorithm1}
\Input{Graph job ${\bm G}^{{\bm A}}$, VC graph ${\bm G}^{{\bm S}}$}
\Output{The optimal candidate ${\bm C^*}$ for distributing ${\bm G}^{{\bm A}}$ over ${\bm G}^{{\bm S}}$.}
// Stage 1  Candidate searching procedure\;
Initialization: ${\bm G}^{{\bm C}}\mathbf{\leftarrow \emptyset }$\;
${\bm G}^{{\bm C}}\mathbf{\leftarrow }\{ {\bm G}^{C_{i}}=\left( {\bm V}^{i}\mathbf{,}{\bm E}^{i},{\bm W}^{i} \right)|i\in \left\{ 1,2,\mathellipsis ,C_{K}^{n} \right\},$ $ \left| {\bm V}^{i} \right|=\left| {\bm V}^{{\bm A}} \right|, {\bm V}^{i}\subseteq {\bm V}^{{\bm S}}, $ ${\bm E}^{i}\subseteq {\bm E}^{{\bm S}}, {\bm W}^{i}\subseteq {\bm W}^{{\bm S}} \};$ \\
${\bm G}^{{\bm C}}\mathbf{\leftarrow }\{ {\bm G}^{C_{j}}\mathbf{|}{\bm E}^{{\bm A}}\subseteq {\bm E}^{j}\mathbf{ }\mbox{~and meet constraints} \left( c \right),\left( d \right)$ $ {\rm in} ~(1) \};$ \\
// Stage 2 The optimal candidate selection procedure\\
\For{each candidate in ${\bm G}^{{\bm C}}$}
{calculate the value of (4);}
The optimal candidate ${\bm C^*}\leftarrow$ the candidate in ${\bm G}^{\bm C}$ with the minimum value of (4);\\
\textbf{End}
\caption{The optimal graph job allocation.}
\end{algorithm}

\begin{algorithm*}[!t]
\Input{Graph job ${\bm G}^{{\bm A}}$, VC graph ${\bm G}^{{\bm S}}$, number of iterations $r$.}
\Output{The applicable candidate ${\bm C^*}$ for distributing ${\bm G}^{{\bm A}}$ over ${\bm G}^{{\bm S}}$.}
// Initialization\;
\For{$s_{j}\in {\bm M }$ }
{$count{\_s}_{j}\leftarrow 0;\gamma_{j}^{Trans}\leftarrow t_{jm}^{Trans};$}
$\tau^{C_{0}}=+\infty ;
\mathrm{\mathbf{\Theta }}\leftarrow \emptyset ;
\mathrm{ }\mathrm{\mathbf{\Theta }}^{{\bm S'}}\leftarrow \emptyset ;
\mathrm{\mathbf{\Theta }}^{{\bm S}}\leftarrow \emptyset ;
{\bm C^*\leftarrow \emptyset };$ $
\mathrm{ }{\bm C}_{\mathbf{\textbf{\textit{Iteration}}}}\mathbf{\leftarrow \emptyset }$\;
// Procedure of the proposed randomized graph job allocation via hierarchical tree based subgraph isomorphism\;
\For{$\textbf{\textit{Iteration}}=1$ to $r$}
{$k\leftarrow 1,{\bm V}^{{\bm A'}}\leftarrow \emptyset $\;
Slot available $\leftarrow true$\;
\While{$k\leq |{\bm N}|$ and Slot available}
{\eIf{there is no idle slots available on any SP}
{Slot available $\leftarrow false$, ${\bm C}_{\textbf{\textit{Iteration}}}\leftarrow \emptyset $}{randomly select one component ${\bm \alpha}_{{\bm k}}\leftarrow $ $\{\upsilon _{i}^{A}|\upsilon_{i}^{A}\in {\bm V}^{\bm A}\}$ from ${\bm G}^{\bm A}$ and place it uniformly at random in one of the free slots on a SP denoted as $s_{j}$\;
$count{\_s}_{j}\leftarrow count{\_s}_{j}+1$\;
$ \gamma_{j}^{Trans}\leftarrow count{\_s}_{j}\cdot t_{jm}^{Trans}$\;
$\bm {C_{\textit{\textbf{Iteration}}}}\leftarrow \{{\{{\bm \alpha }_{{k}},s}_{j}\}\}$\;
${\bm V}^{{\bm A}'}\leftarrow {\bm V}^{{\bm A}}\backslash{\bm \alpha}_{{ k}}$\;
$\mathrm{\mathbf{\Theta }}^{{\bm S}}\mathbf{\leftarrow \{}s_{j}\}$\;
\For{$k\leq \left| {\bm N} \right|\mathrm{ }$}
{~place components in ${\mathrm{\mathbf{\Theta
}}^{{\bm A}}=\{\upsilon_{\mathbf{\bm \alpha
}_{ k_1}}^{A}\mathrm{,\mathellipsis }\upsilon_{\mathbf{\bm \alpha
}_{k_{|\mathbf{\Theta}^{\bm A}|}}}^{A}\mathrm{ \mathbf{\}\in
}}{\bm V}}^{{\bm A'}}$ that have a one-hop
connection with components in ${\bm {\alpha }}_{{k}}$ uniformly at
random in $|\mathbf{\Theta}^{\bm A}|$ free slots denoted as ${\mathrm{\mathbf{\Theta
}}^{{\bm S}'}\mathrm{\mathbf{=\{}}S_{\mathbf{\bm \alpha
}_{{k_1}}}\mathrm{,\mathellipsis }, S_{\mathbf{\bm \alpha}_{{\bm k_{|\mathbf{\Theta}^{\bm A}|}}}}\}
}$ by satisfying \textbf{(c-1)} and \textbf{(c-2) }and\textbf{ (c-3)}\;

\textbf{(c-1)}\:
\For{$b=1$ to $|\mathbf{\Theta}^{\bm A}|$}
{ \quad $e^{{- \gamma_{\mathbf{\bm \alpha }_{{k_b}}}^{Trans}}\lambda_{\mathbf{\bm \alpha }_{{k_{bm}}}}}\geq \xi $}

\textbf{(c-2)}\:
\If{edge $e_{zz'}^{A}\in {\bm E}^{{\bm A}}$, where $z\in {\bm{\Theta }}$ and $z'\in {\bm\Theta }$, $z$ and $z'$ are allocated to two SPs $l$ and $l'$ \\ \qquad~  respectively, ${l,l'\in {\bm \Theta }}^{S'}$ and $l\ne l'$,}
{\quad  there must be an edge $e_{ll'}^{S}\in {\bm E}^{{\bm S}}$ and satisfying: ${e}^{-(|t_{lm}^{Trans}-t_{l'm}^{Trans}|+\omega_{zz'}^{A})\lambda_{ll'}}\geq \varepsilon $}

\textbf{(c-3)}\:
\If{edge $e_{zz'}^{A}\in {\bm E}^{{\bm A}}$, where $z\in {\bm{{\alpha}_{k}}}$ and $z'\in {\bm{{\Theta}^{A}}}$, $z$ and $z'$ are allocated to two SPs $l$ and $l'$\\\qquad~   respectively, ${l\in {\bm \Theta }}^{S}$, ${l'\in{\bm\Theta}}^{S'}$ and $l\ne l'$,}
{\quad there must be an edge $e_{ll'}^{S}\in {\bm E}^{{\bm S}}$ and satisfying: ${e}^{-(|t_{lm}^{Trans}-t_{l'm}^{Trans}|+\omega_{zz'}^{A})\lambda_{ll'}}\geq \varepsilon $}

$k=k+|\mathbf{\Theta}^{\bm A}|$\;
$\mathrm{\mathbf{\Theta }}^{{\bm S}}\leftarrow \mathrm{\mathbf{\Theta }}^{{\bm S}'};$\\
\For{$b=1$ to $|\mathbf{\Theta}^{\bm A}|$}
{$count\_s_{{\bm \alpha }_{{k_b}}}=count\_s_{{\bm \alpha }_{{k_b}}}+1$\;
$\gamma_{s_{\mathbf{\bm \alpha }_{{ k_b}}}}^{Trans}\leftarrow count\_s_{\mathbf{\bm  \alpha }_{{k_b}}}\cdot t_{s_{\mathbf{\bm  \alpha }_{{k_{bm}}}}}^{Trans};$ }
$\bm C_{\textit{\textbf{Iteration}}}\leftarrow \bm C_{\textit{\textbf{Iteration}}}\cup \{\mathrm{\mathbf{\Theta}}^{{\bm A}},\mathrm{\mathbf{\Theta }}^{{\bm S}'}\}$\;
${\bm V}^{{\bm A'}}\leftarrow {\bm V}^{{\bm A}'}\backslash\mathrm{\mathbf{\Theta }}^{{\bm A}}$\;
${\bm \alpha }_{{k}}\leftarrow \mathrm{\mathbf{\Theta }}^{{\bm A}};$
}
}
}
$\tau^{C_{\textit{\textbf{Iteration}}}}\leftarrow \mathrm{the~value~of~}\left(\mathrm{4} \right)$ on candidate $\textbf{\textit{C}}_{\bm {Iteration}}$\;
\For{$\tau^{C_{\textit{Iteration}}}>\tau^{C_{\textit{Iteration}-1}}$}
{${\bm C}_{\bm {Iteration}}\leftarrow {\bm C}_{\bm {Iteration-1}};$}
}
${\bm C^*}\leftarrow {\bm C}_{\bm {Iteration}};$\\
\textbf{End}

\caption{Randomized graph job allocation via hierarchical tree based subgraph isomorphism.}
\end{algorithm*}

Given that the objective function in (4) represents an NIP problem that is
NP-hard, a JO can rarely identify a solution to reconfigure the IoV
extemporaneously, as the running time needed to solve large and real-life
network cases increases sharply in line with vehicular density.
To solve the graph job allocation problem, we first propose an optimal
method for low-traffic conditions (e.g., fewer than
five SPs in a VC). Afterward, in rush-hour scenarios, we design a randomized
graph job allocation mechanism via hierarchical tree based subgraph isomorphism, through which near-optimal
solutions can be obtained with low computational complexity.

\section{Optimal Graph Job Allocation Mechanism}

\begin{figure*}[!t]\centering	
    \subfigure[]{{\includegraphics[width=1.75in]{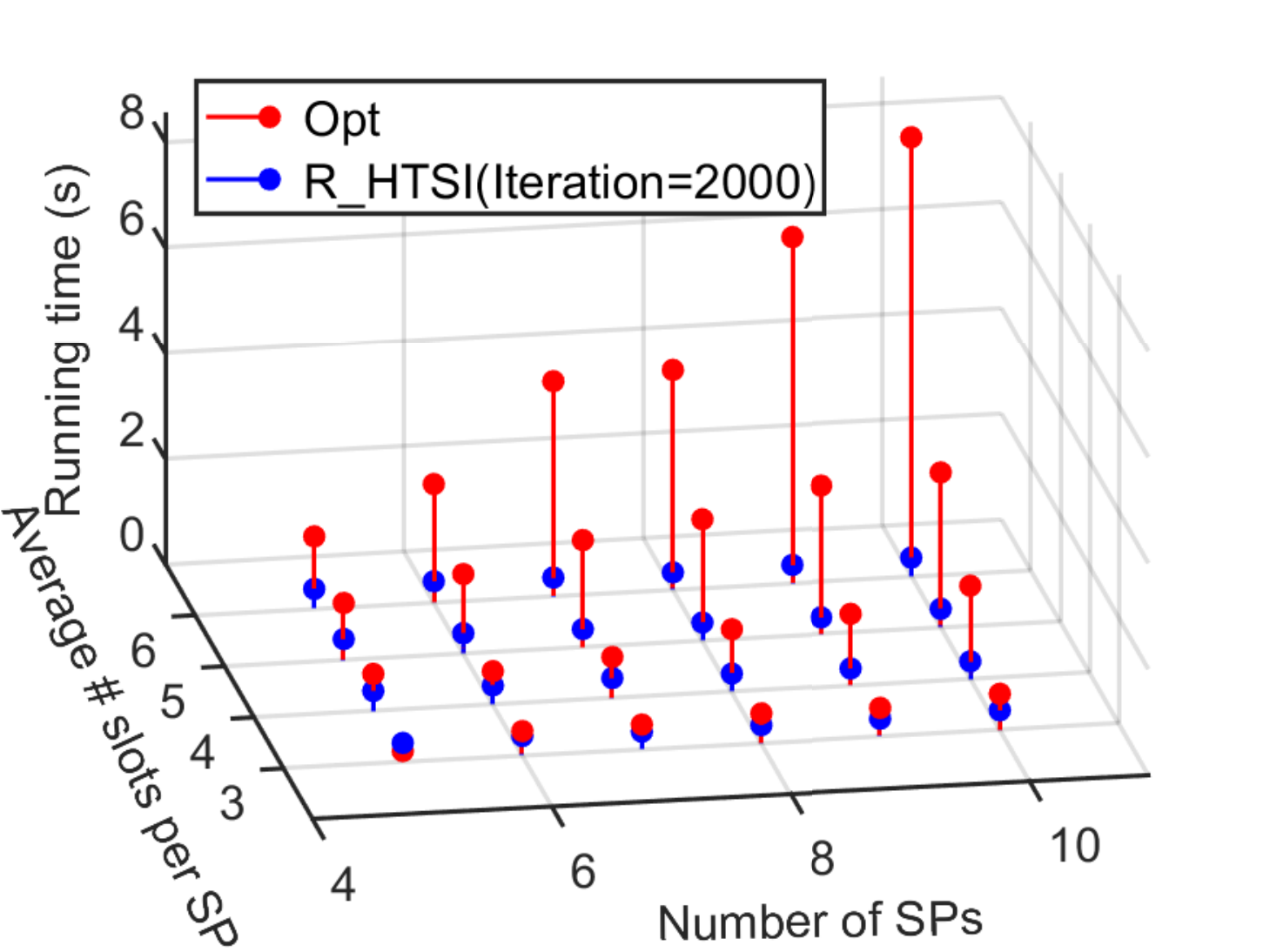}}\hfill
    {\includegraphics[width=1.75in]{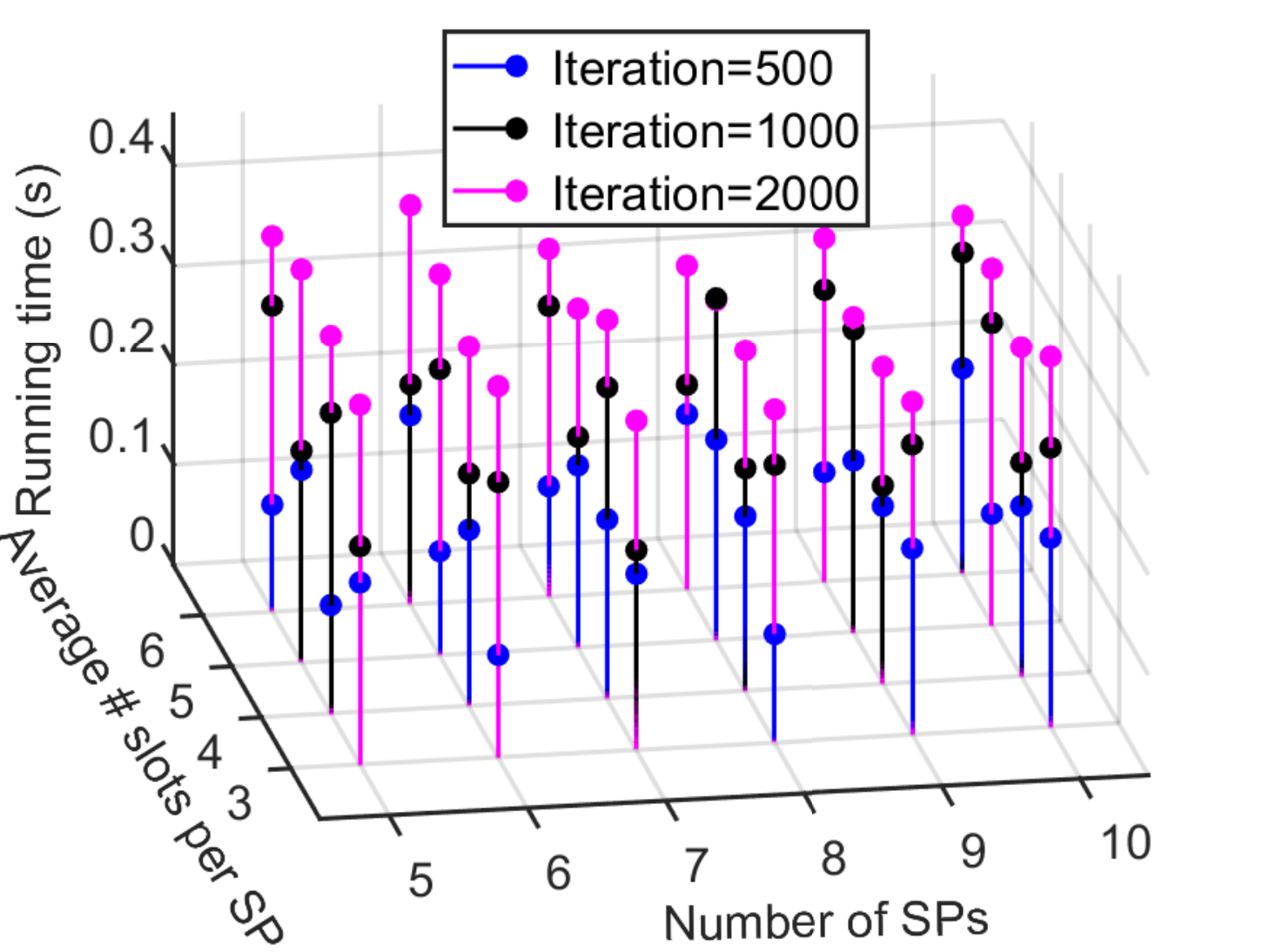}}}\hfill
    \subfigure[]{{\includegraphics[width=1.75in]{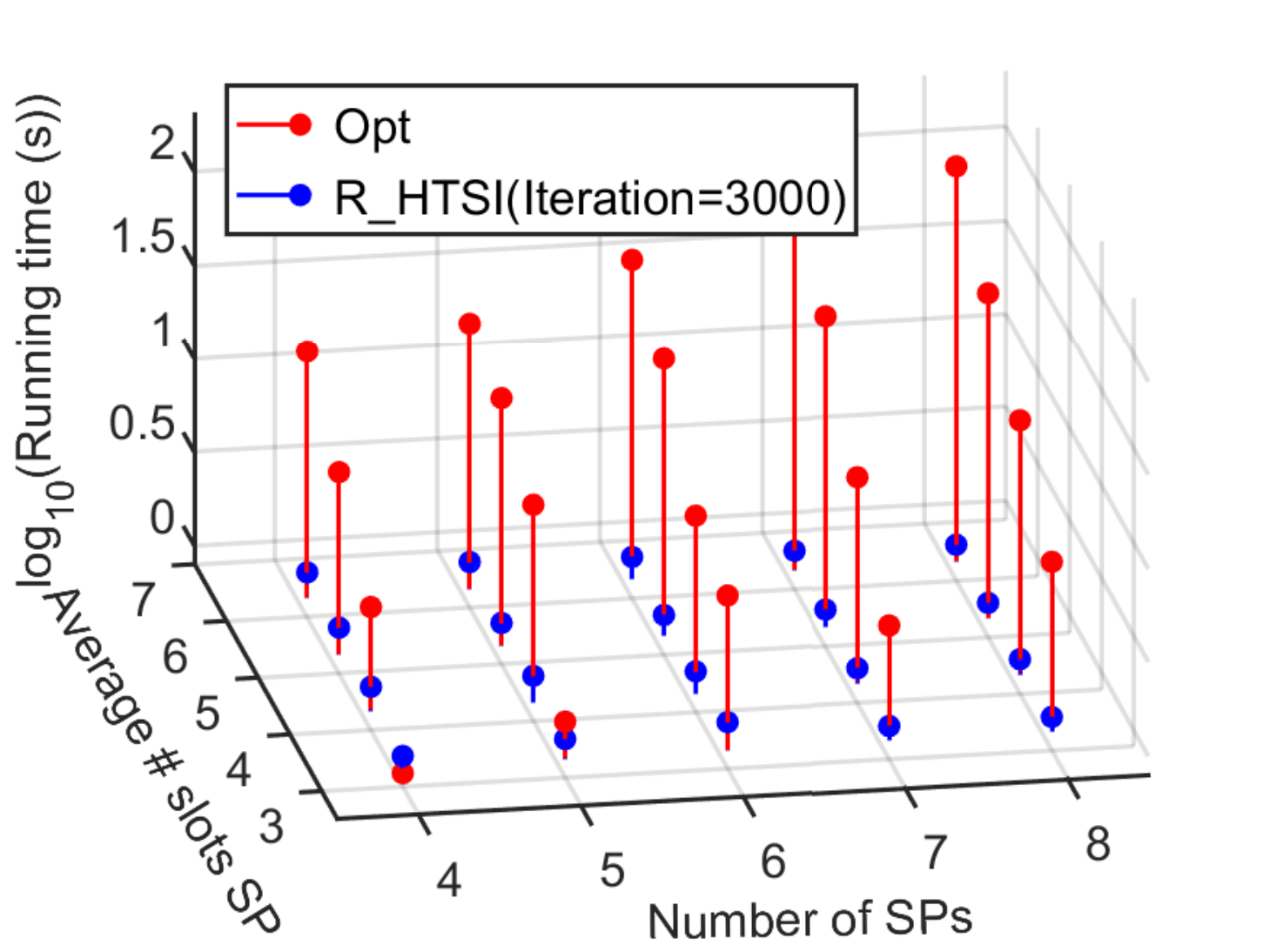}}\hfill
    {\includegraphics[width=1.75in]{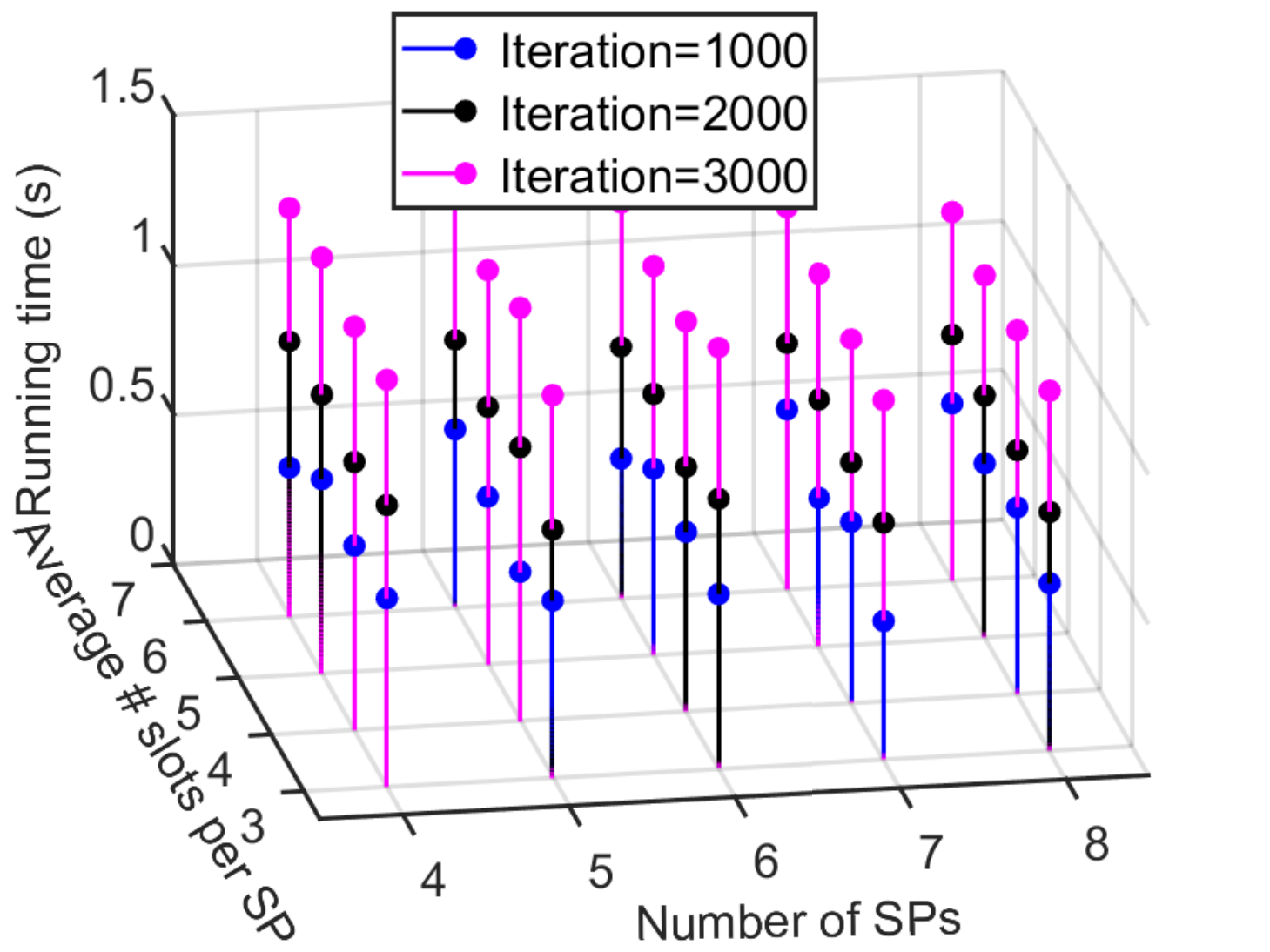}}}\\
    \subfigure[]{{\includegraphics[width=1.75in]{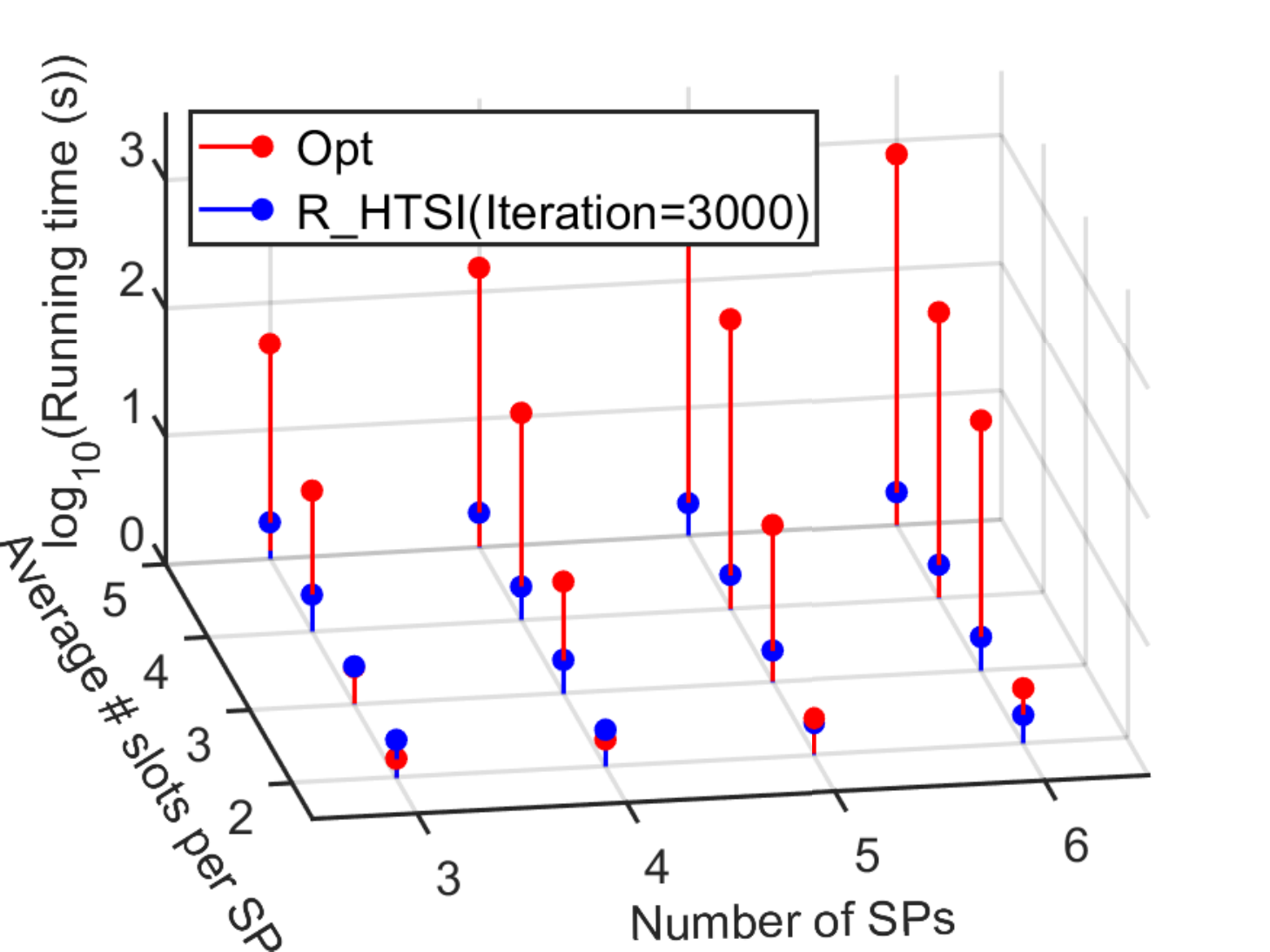}}\hfill
    {\includegraphics[width=1.75in]{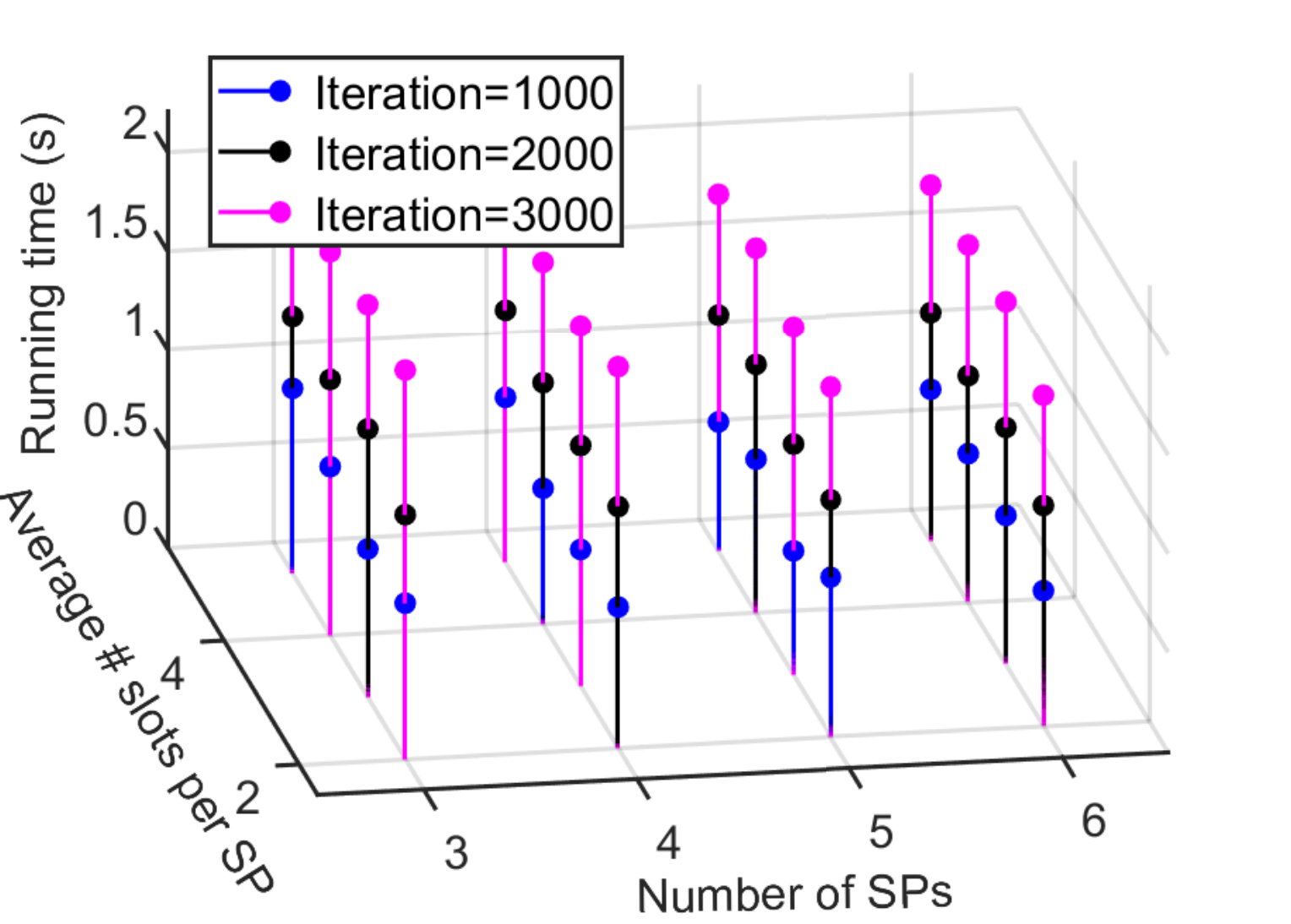}}}\hfill
    \subfigure[]{{\includegraphics[width=1.75in]{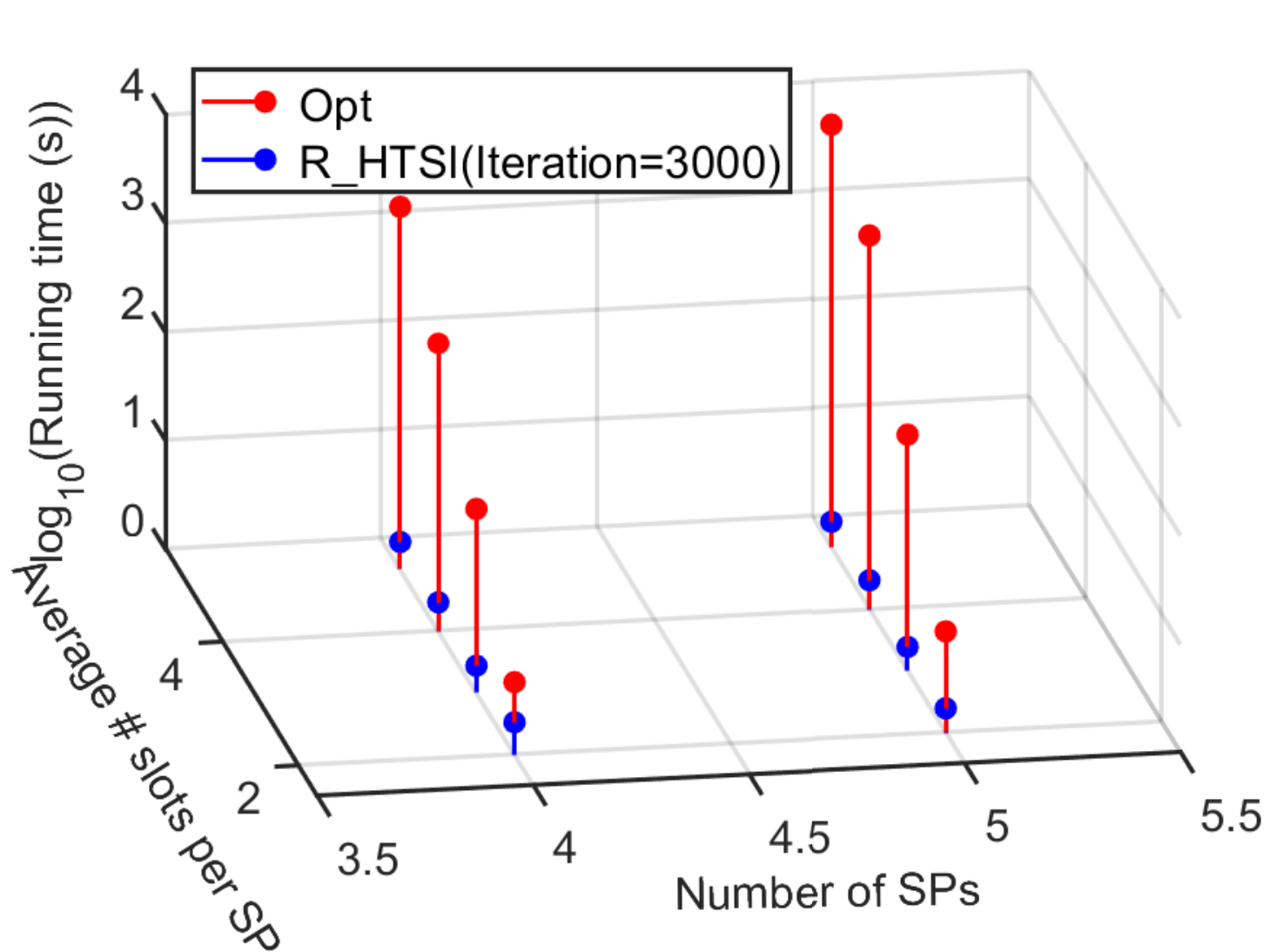}}\hfill
    {\includegraphics[width=1.75in]{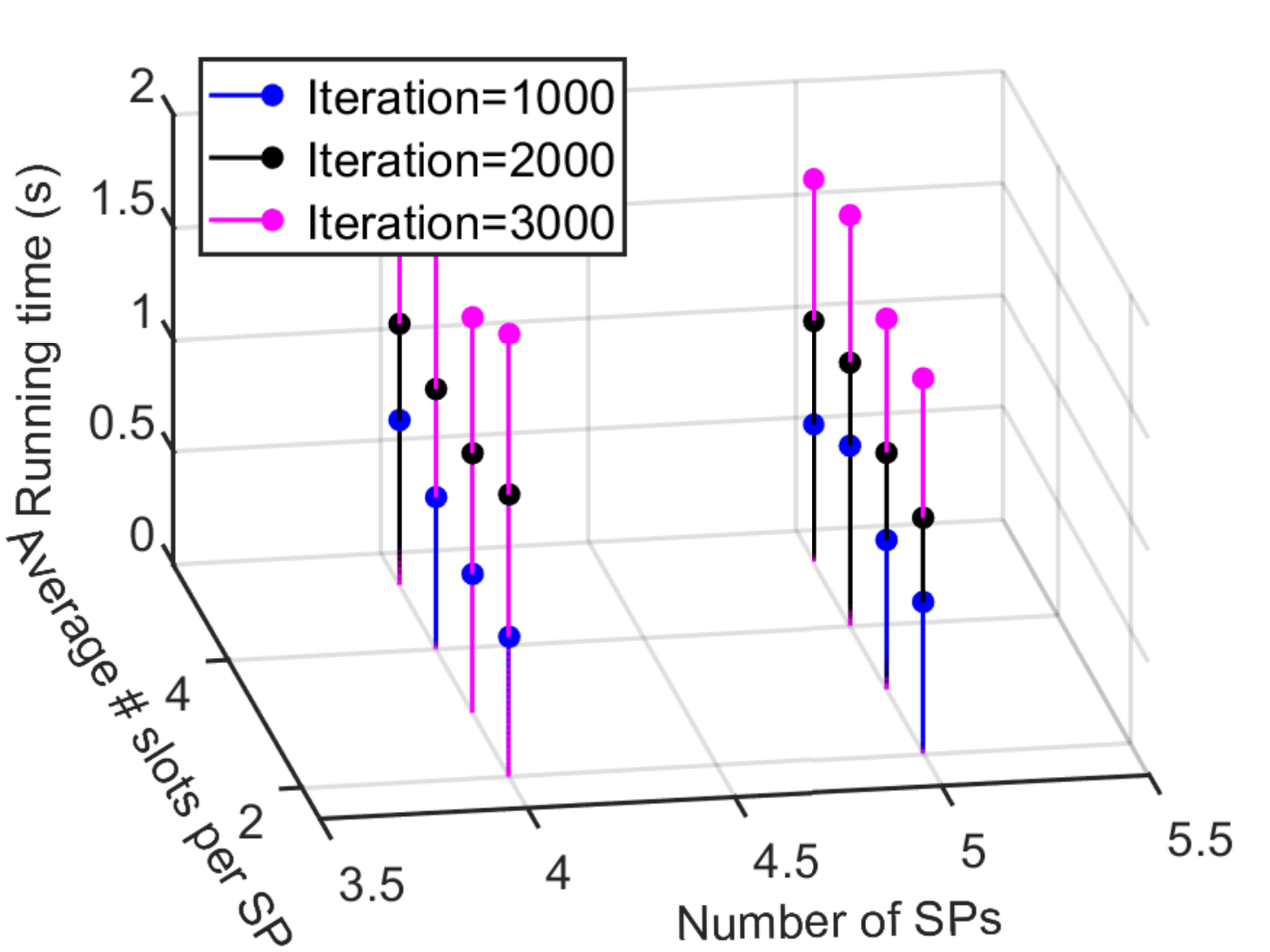}}}\\
    \subfigure[]{{\includegraphics[width=1.75in]{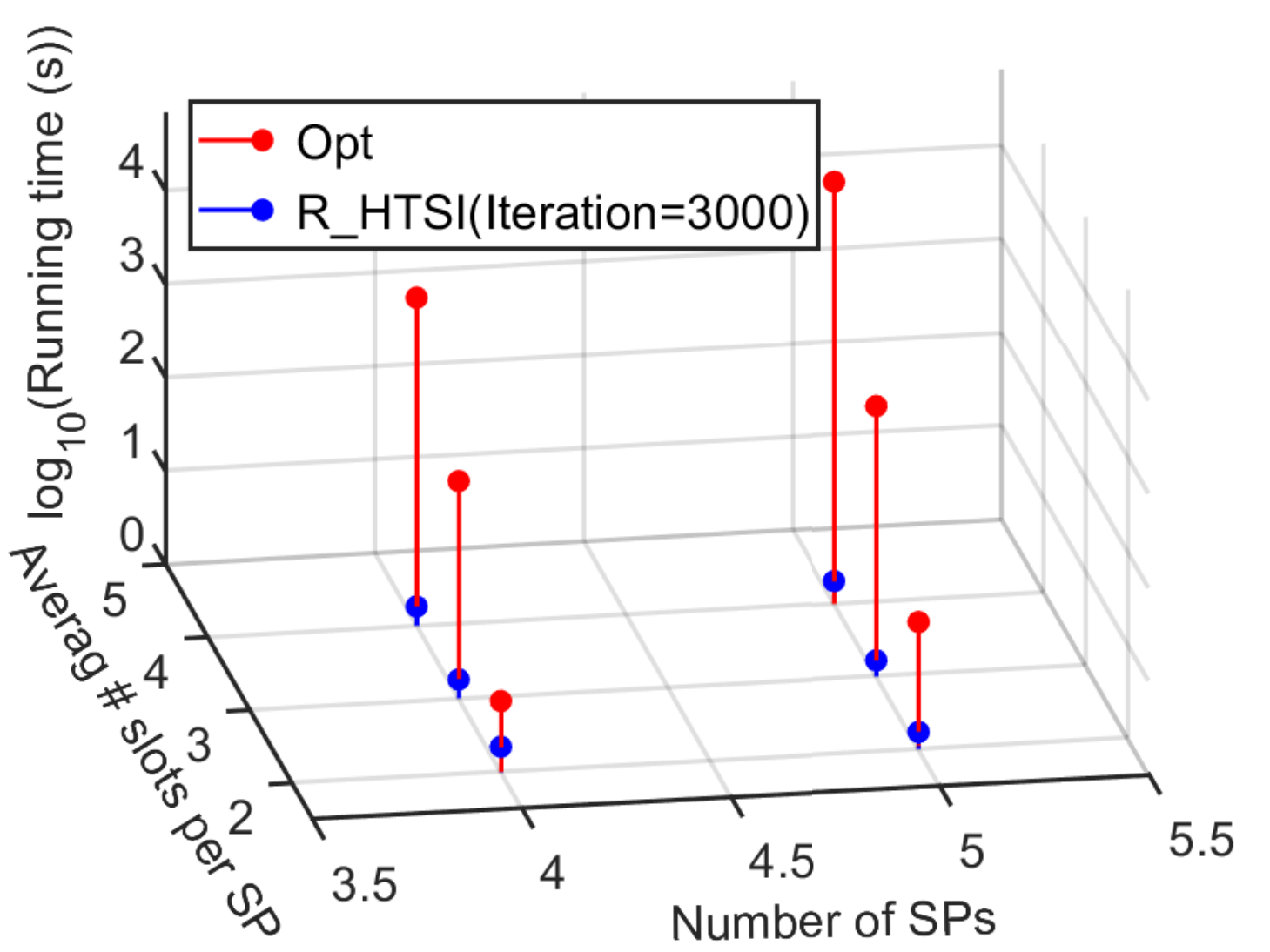}}\hfill
    {\includegraphics[width=1.75in]{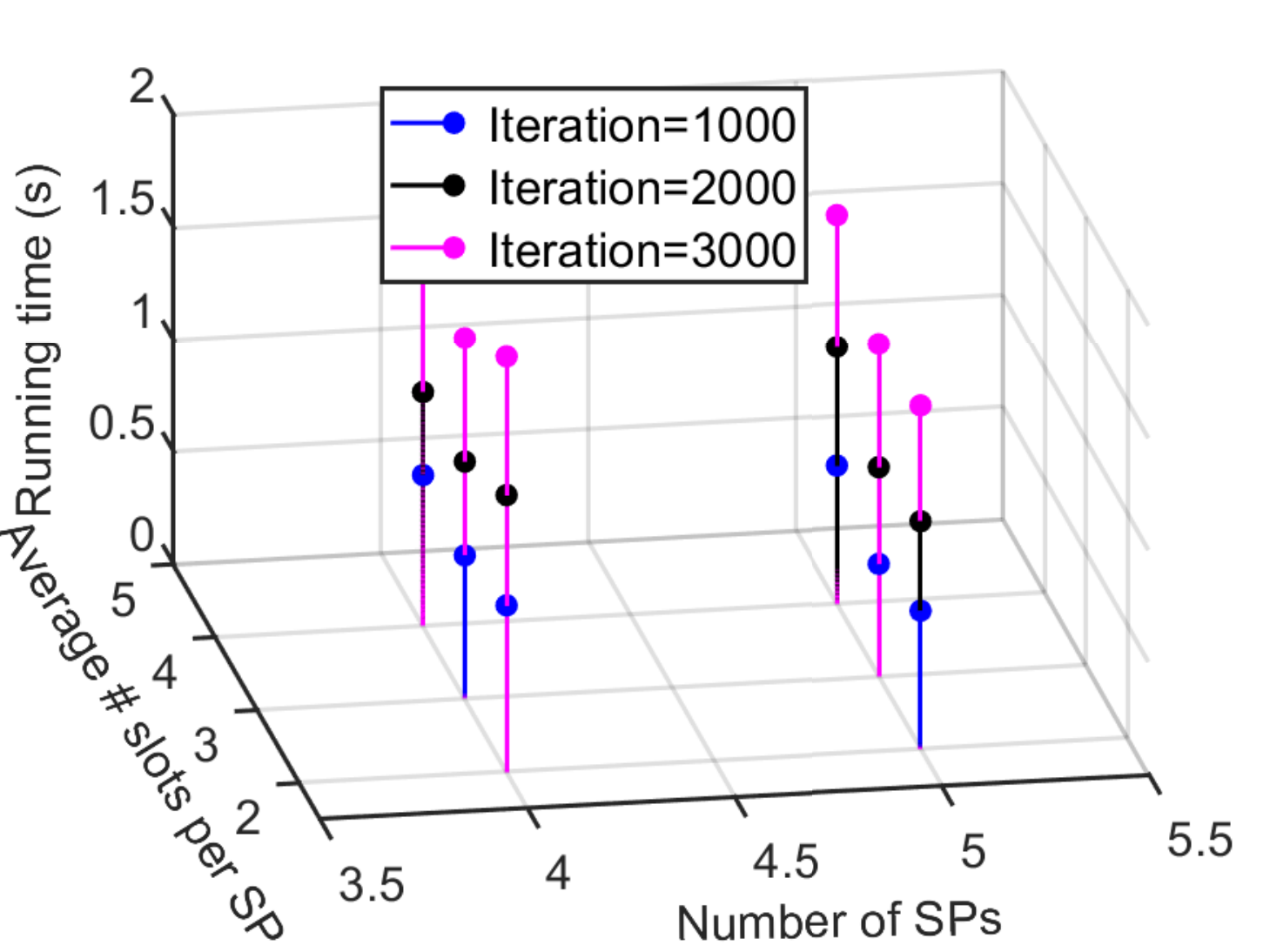}}}\\
\caption{Running time performance: the comparison between the optimal and the
randomized graph job allocation via hierarchical tree based subraph isomorphism mechanisms (left sub-plot in each sub-figure) and the comparison on randomized mechanism with different number of iterations (right sub-plot in each sub-figure): a) running time comparisons of closed triad graph
jobs; b) running time comparisons of square graph jobs; c) running time
comparisons of bull graph jobs; d) running time comparisons of double-star
graph jobs; e) running time comparisons of tadpole graph jobs.}
\label{fig3}
\end{figure*}

\begin{figure*}[!t]\centering	
    \subfigure[]{\includegraphics[width=2.2in]{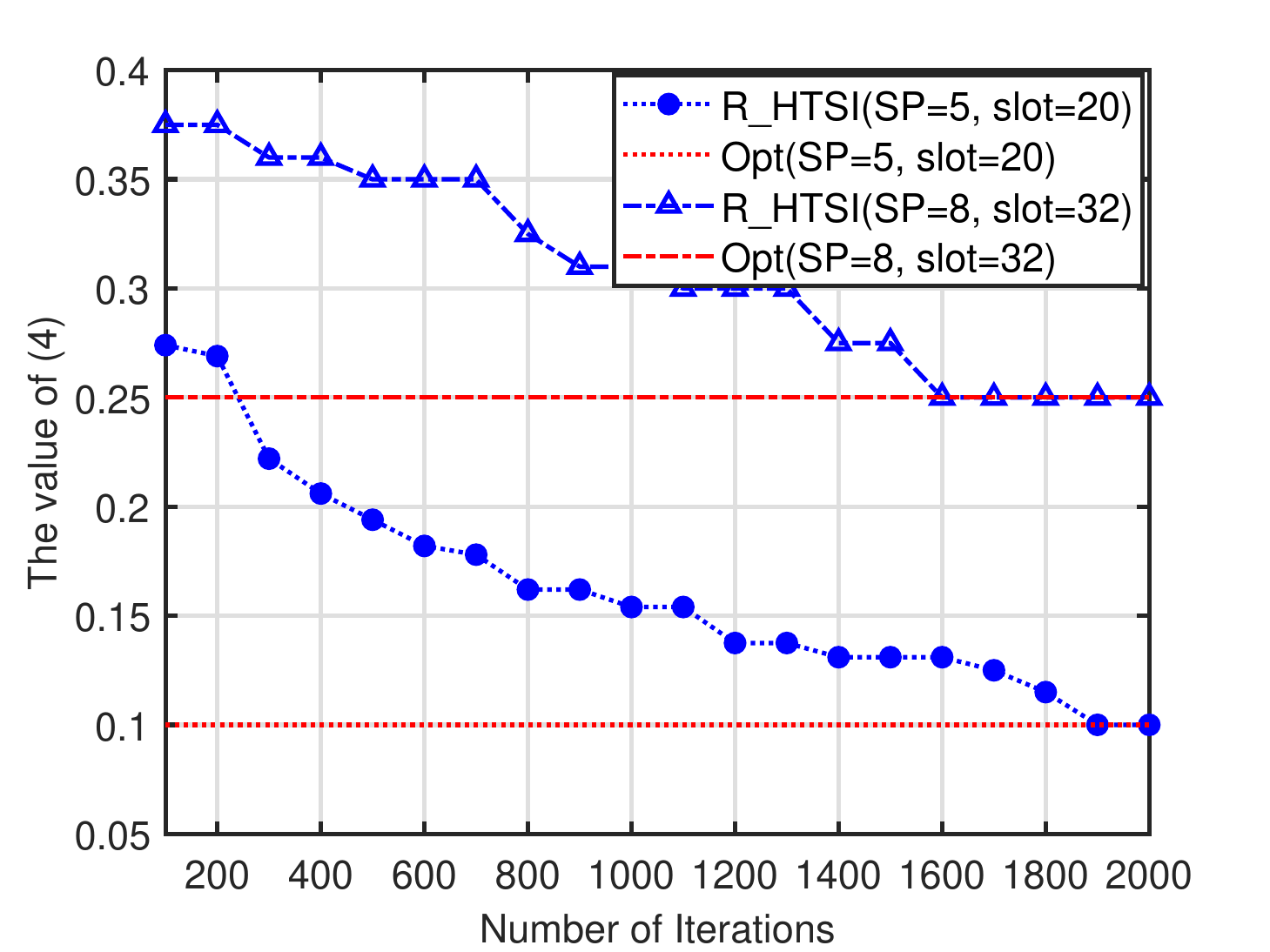}}\quad
    \subfigure[]{\includegraphics[width=2.2in]{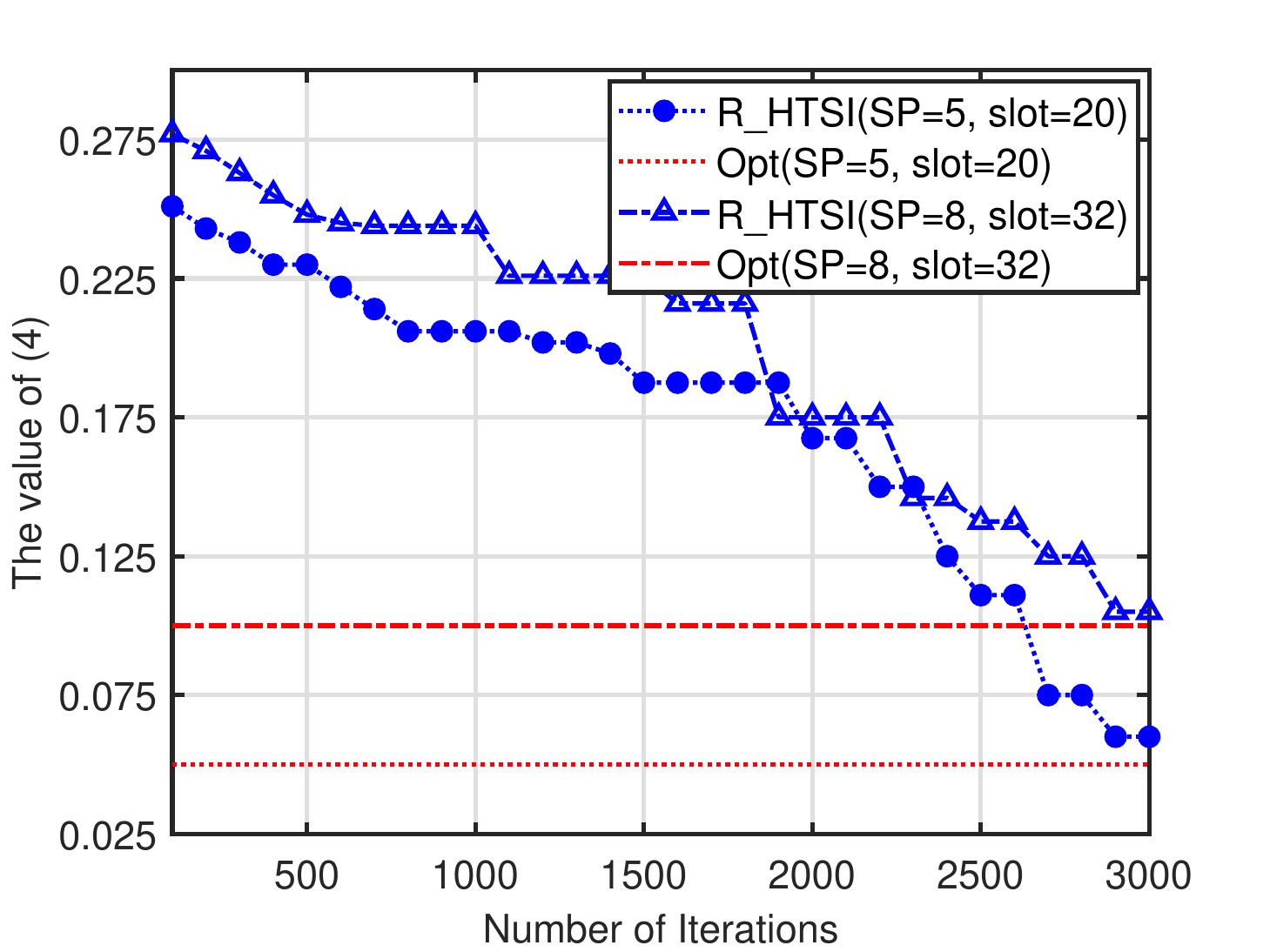}}\quad
    \subfigure[]{\includegraphics[width=2.2in]{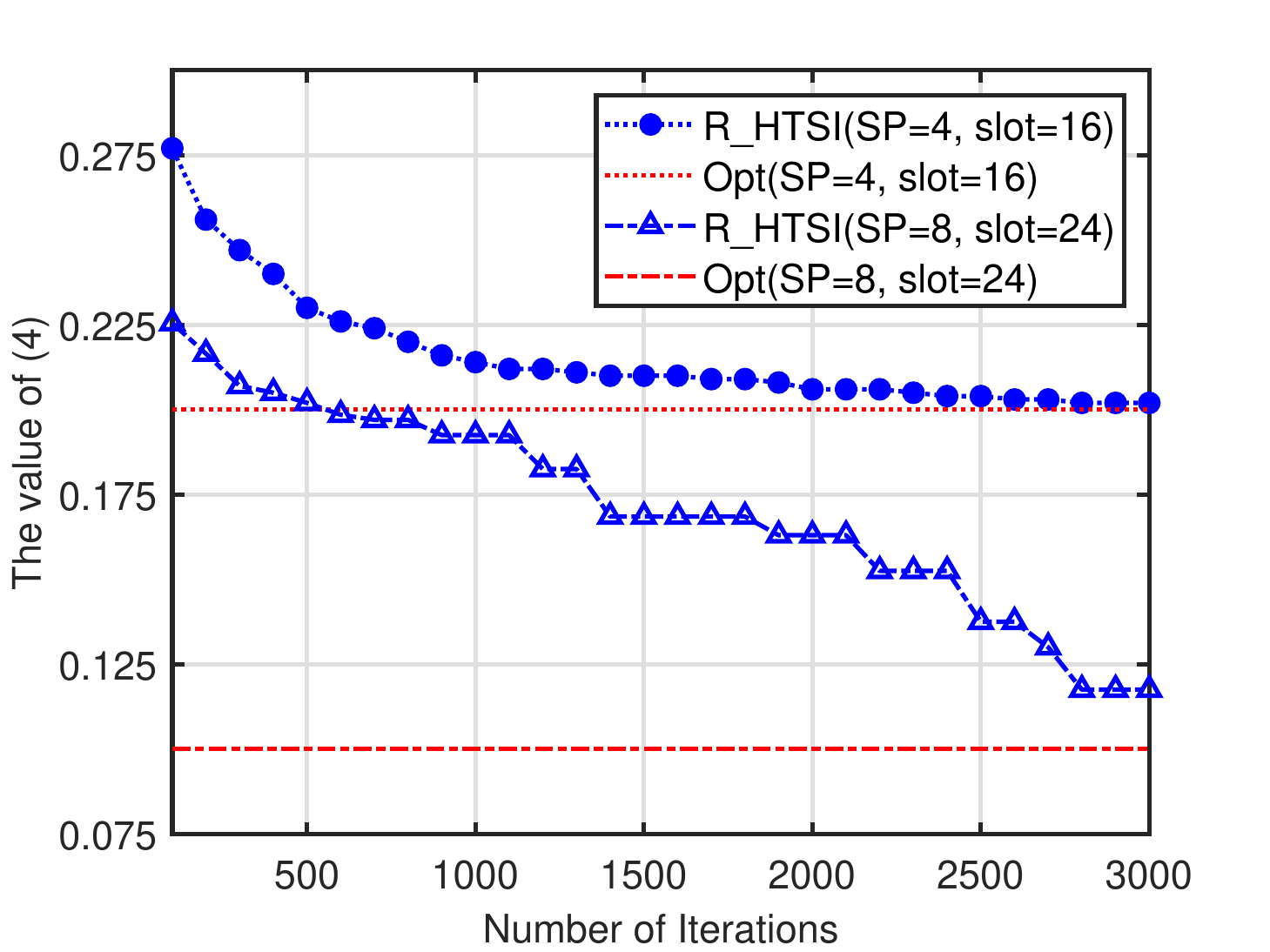}}\\
    \subfigure[]{\includegraphics[width=2.2in]{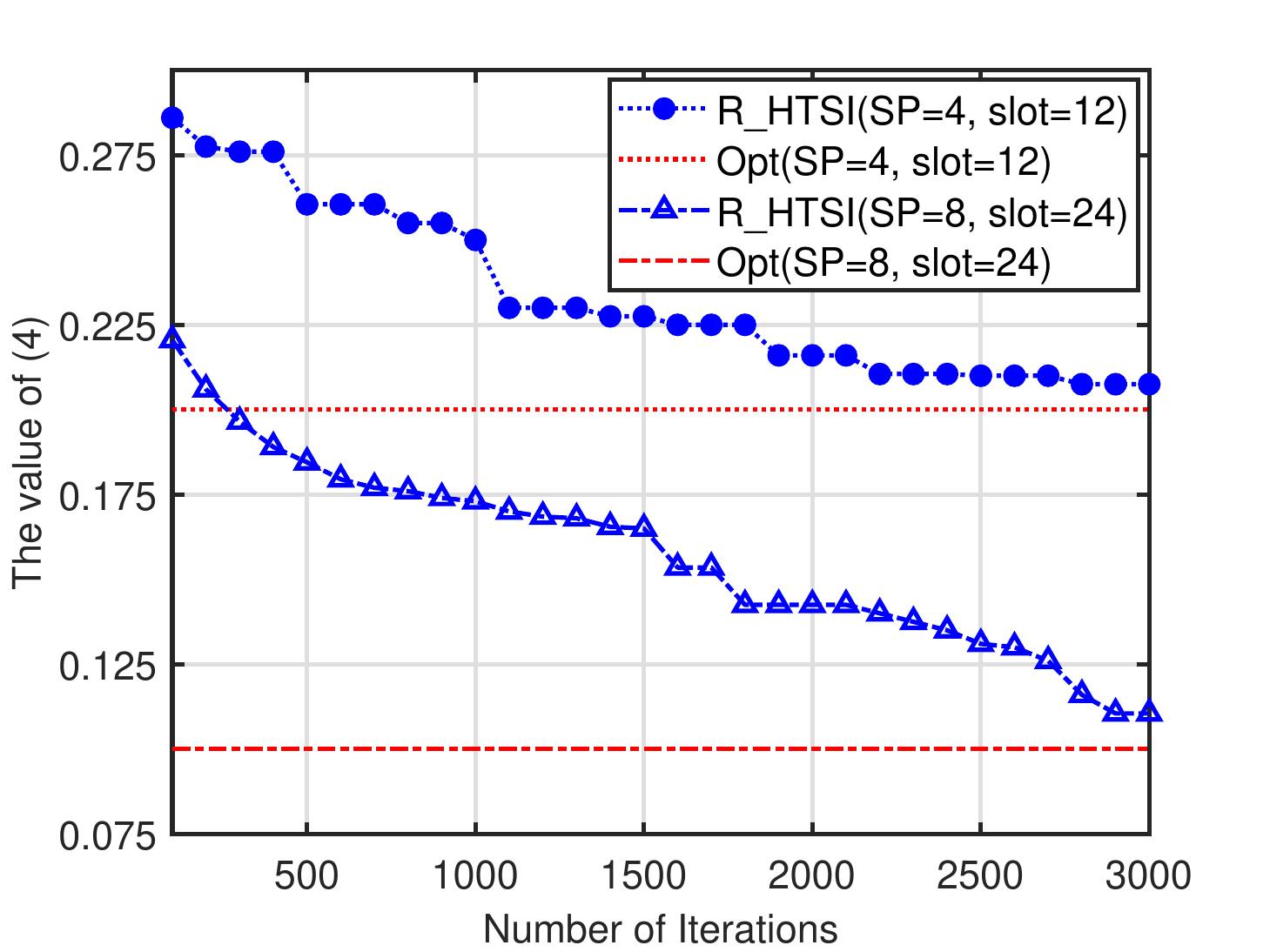}}\qquad
    \subfigure[]{\includegraphics[width=2.2in]{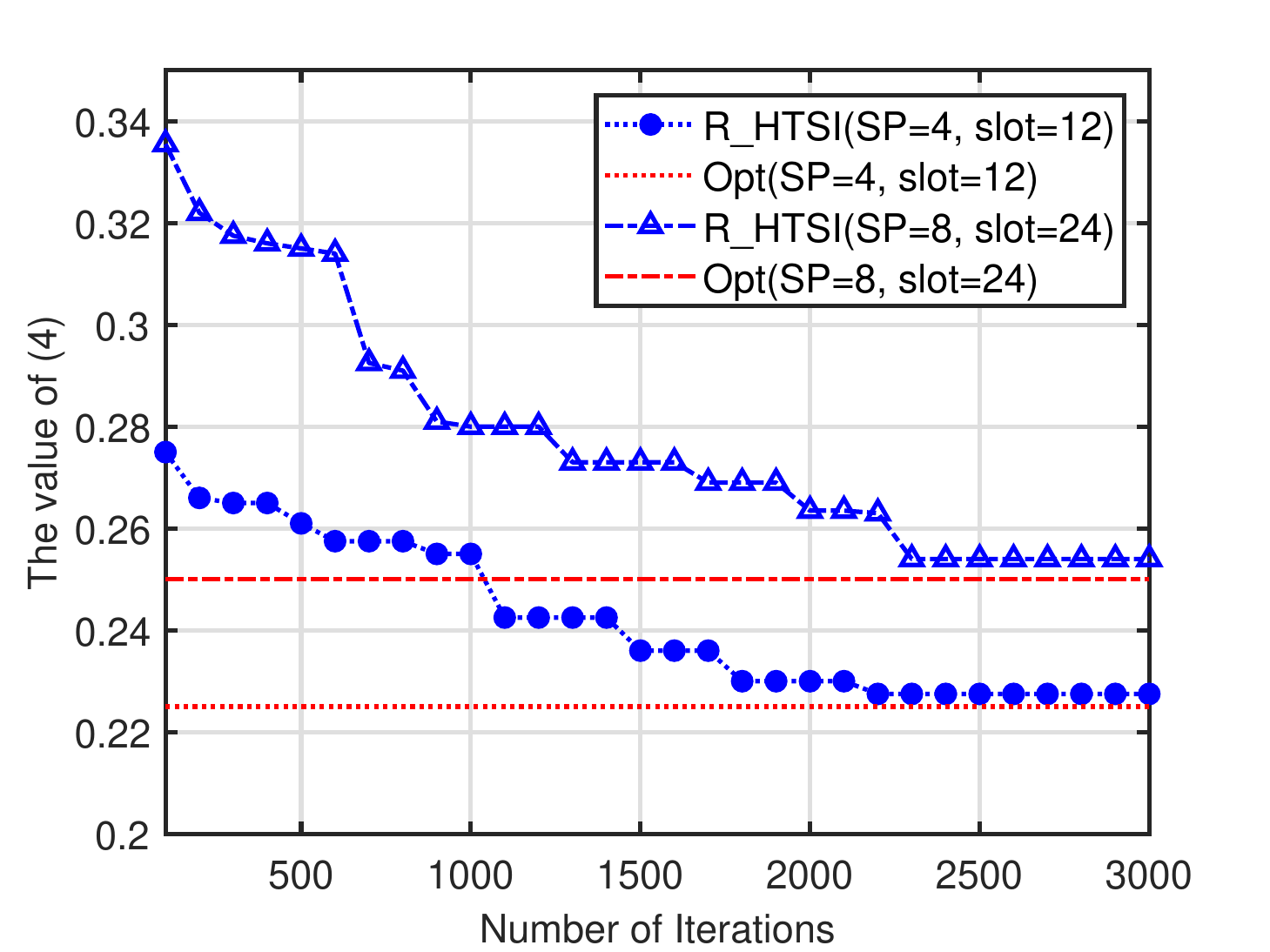}}
\caption{Performance comparison of objective function between optimal and the
randomized graph job allocation via hierarchical tree based subraph isomorphism mechanisms: a) comparison of closed triad graph jobs; b)
comparison of square graph jobs; c) comparison of bull graph jobs; d)
comparison of double-star graph jobs; e) comparison of tadpole graph jobs.}
\label{fig4}
\end{figure*}

\begin{figure*}[!t]\centering	
    \subfigure[]{\includegraphics[width=2.28in]{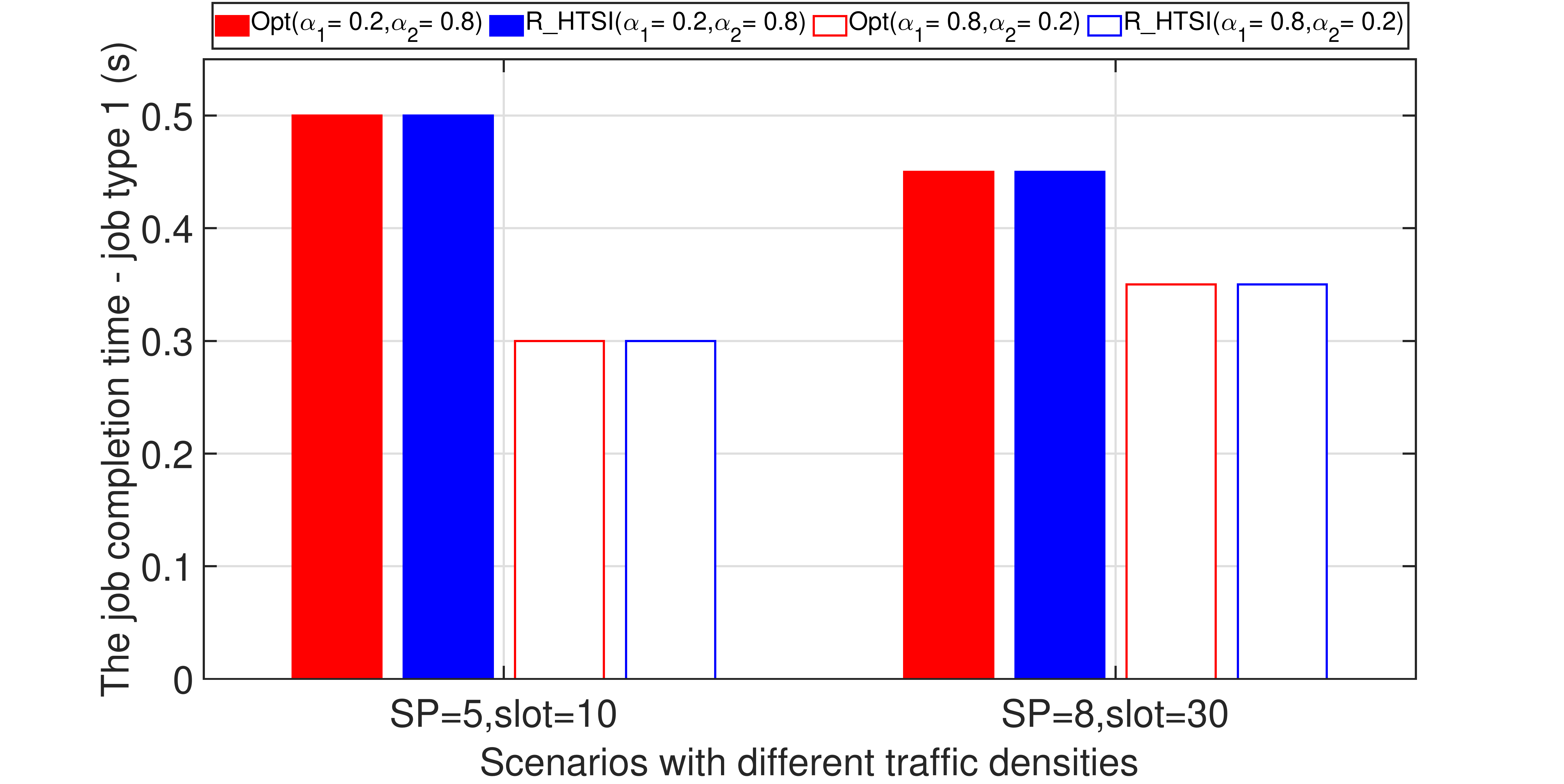}}\quad
    \subfigure[]{\includegraphics[width=2.28in]{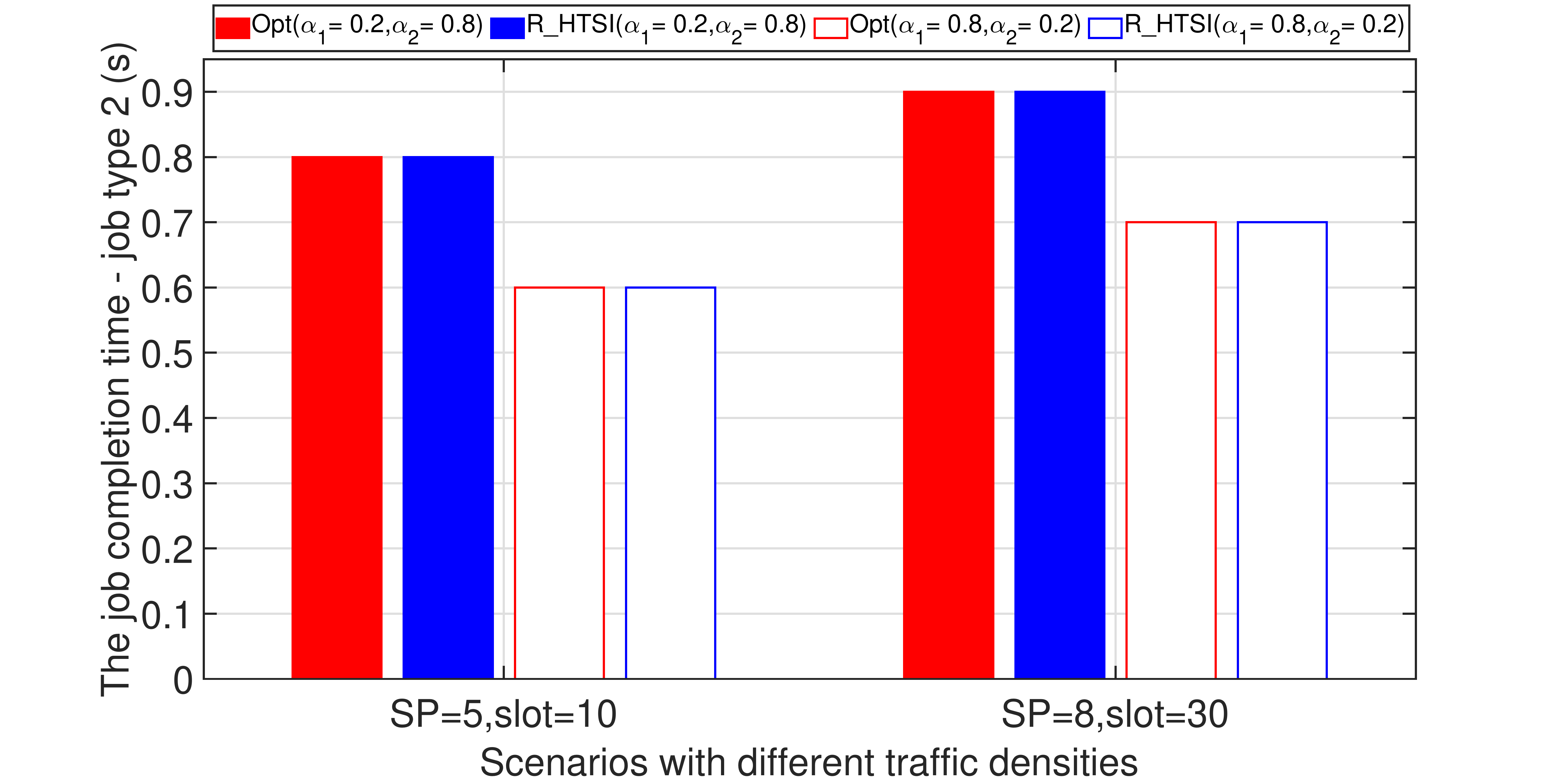}}\quad
    \subfigure[]{\includegraphics[width=2.28in]{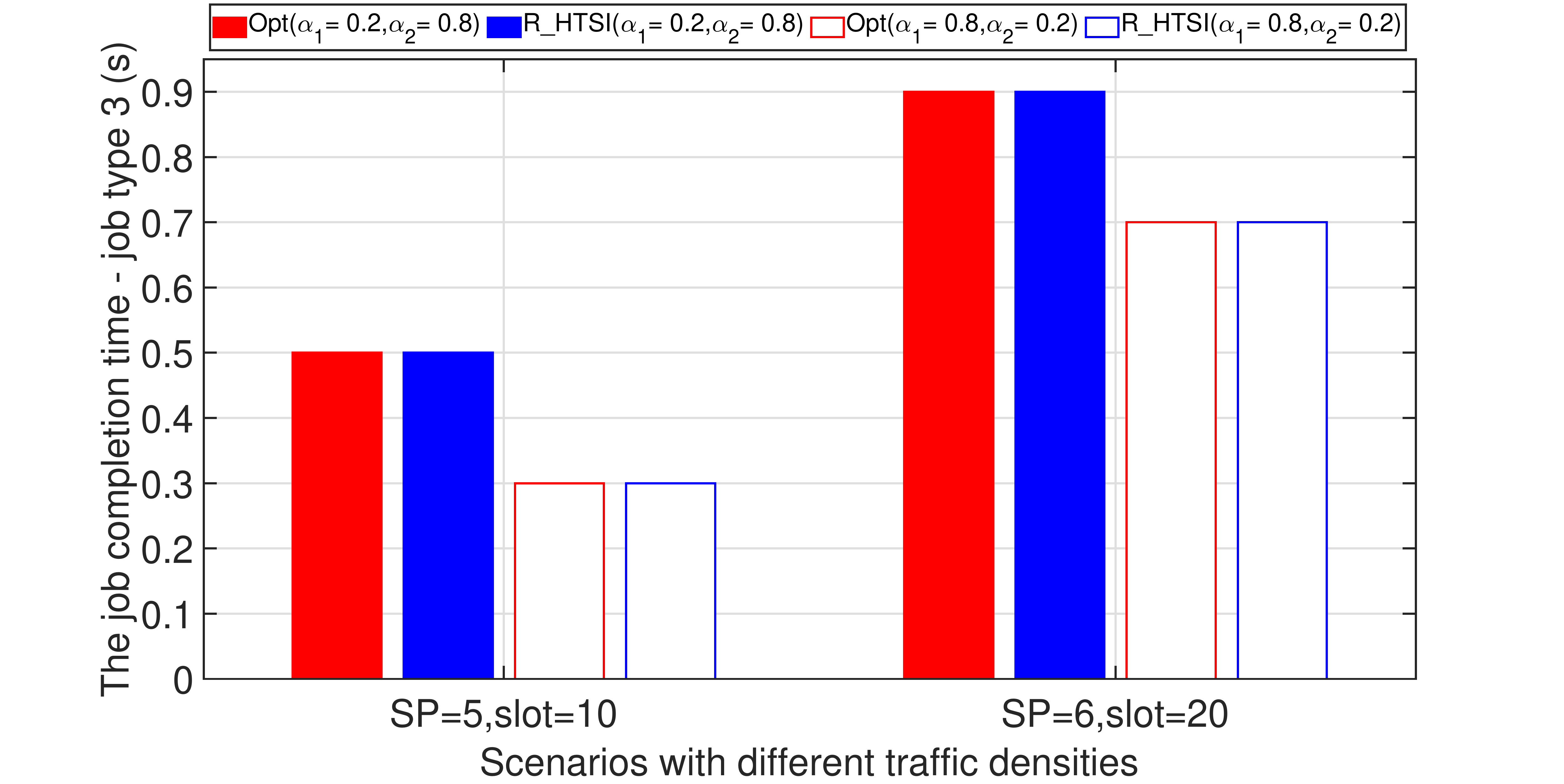}}\\
    \subfigure[]{\includegraphics[width=2.28in]{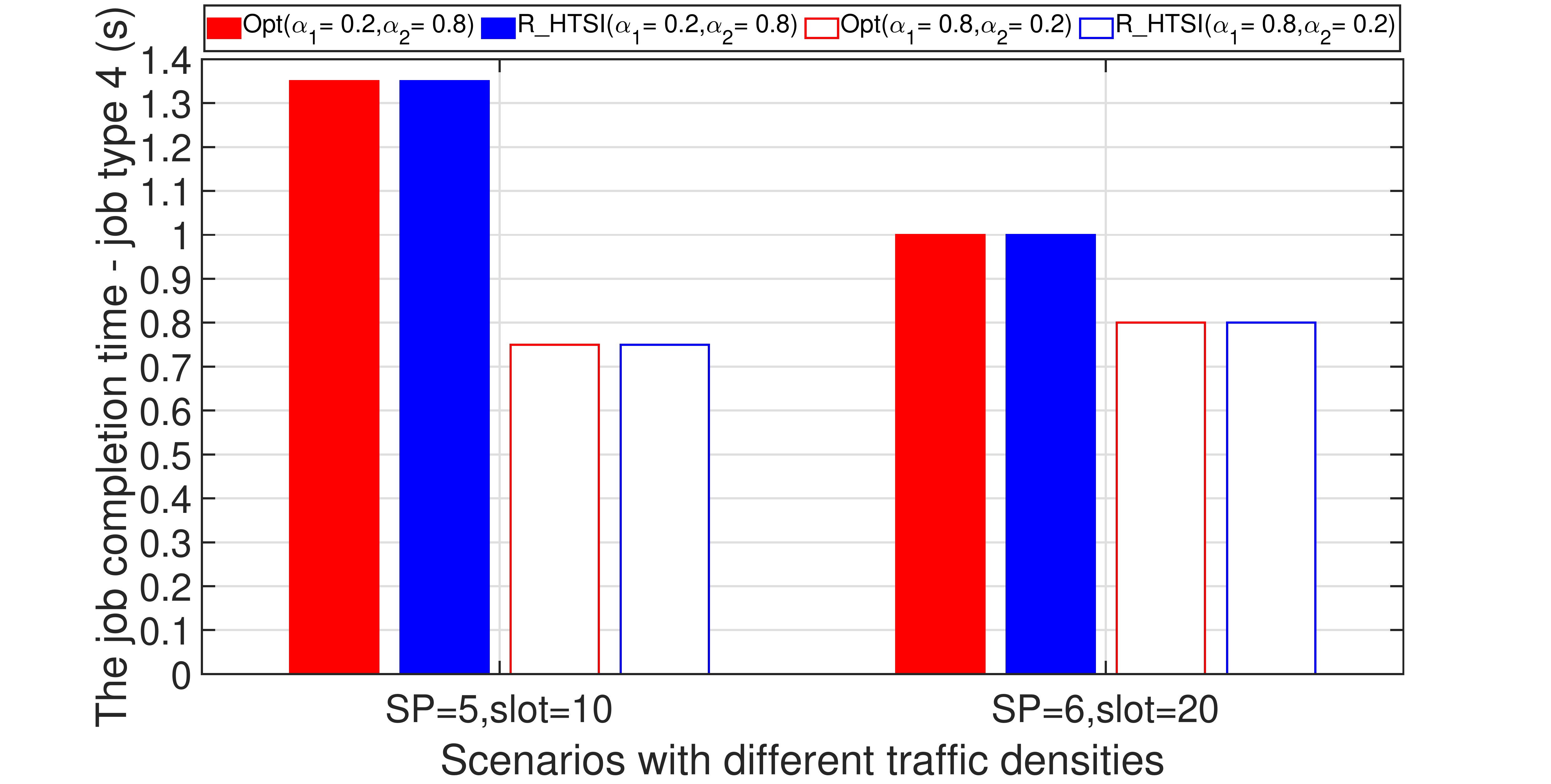}}\qquad
    \subfigure[]{\includegraphics[width=2.28in]{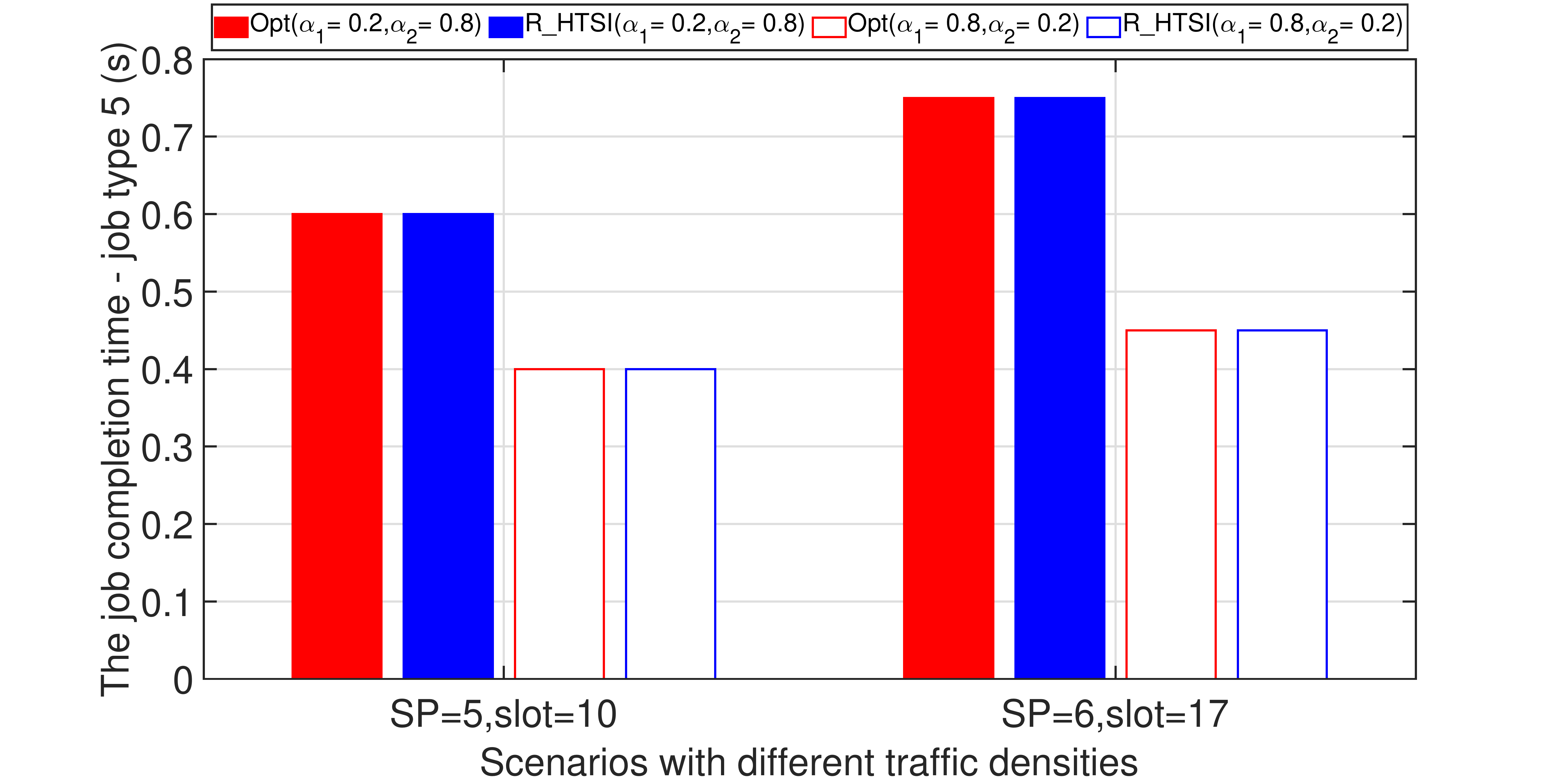}}
\caption{The job completion time upon selection of different values of $\alpha_1$ and $\alpha_2$ in low traffic and rush hour scenarios: a) comparison of closed triad graph jobs; b)
comparison of square graph jobs; c) comparison of bull graph jobs; d)
comparison of double-star graph jobs; e) comparison of tadpole graph jobs.}
\label{fig5}
\end{figure*}

\begin{figure*}[!t]\centering	
    \subfigure[]{\includegraphics[width=2.28in]{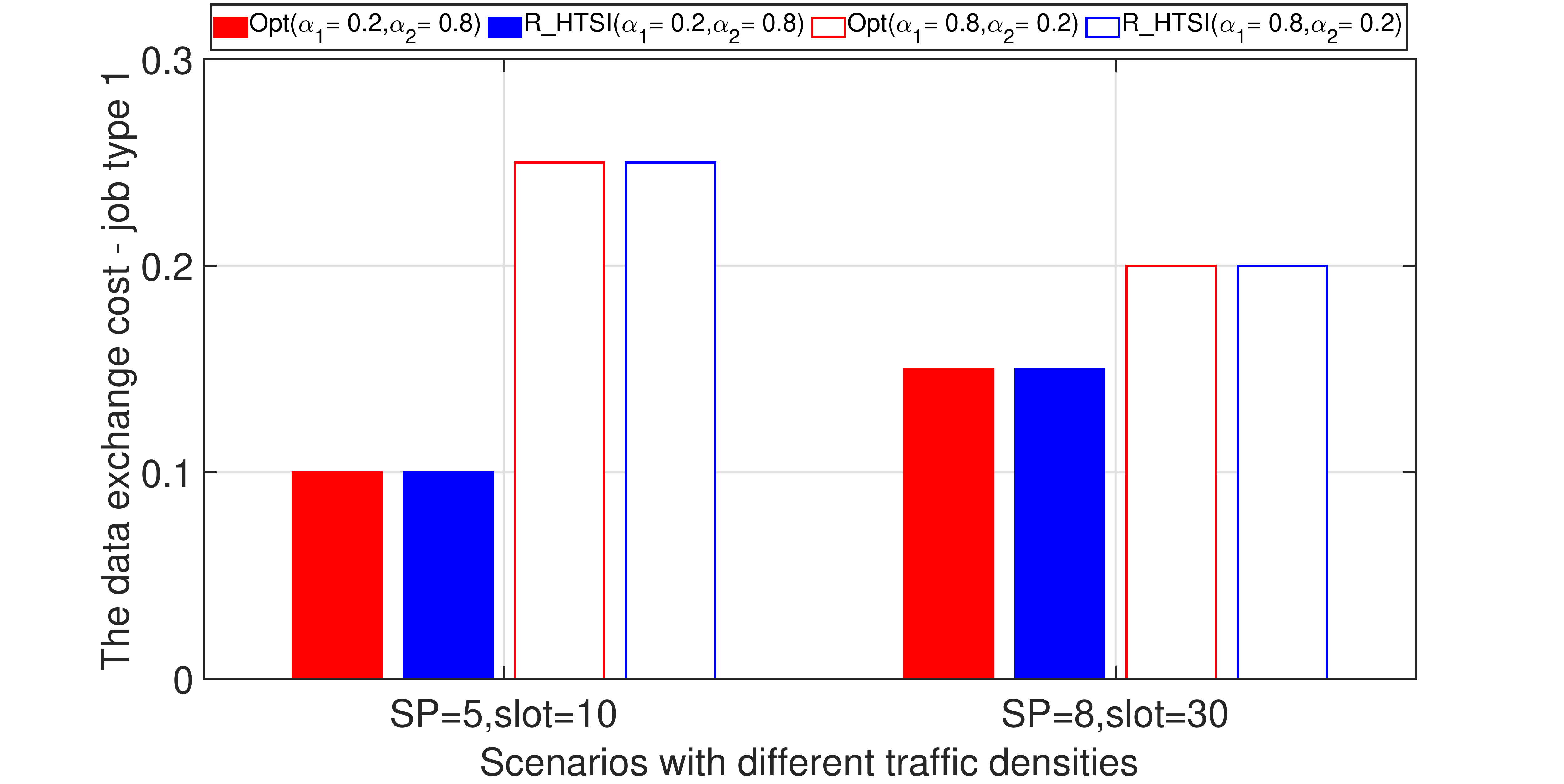}}\quad
    \subfigure[]{\includegraphics[width=2.28in]{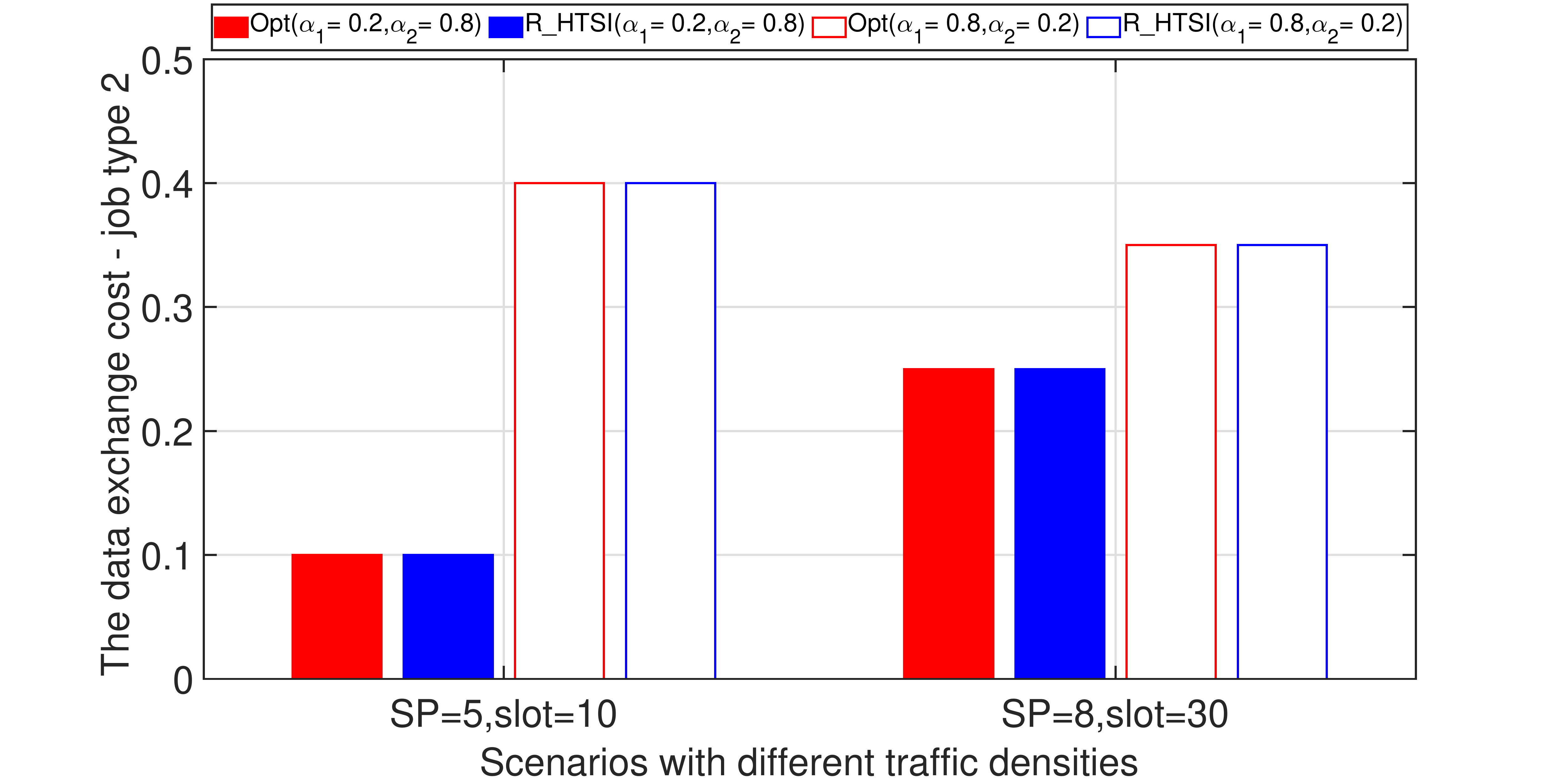}}\quad
    \subfigure[]{\includegraphics[width=2.28in]{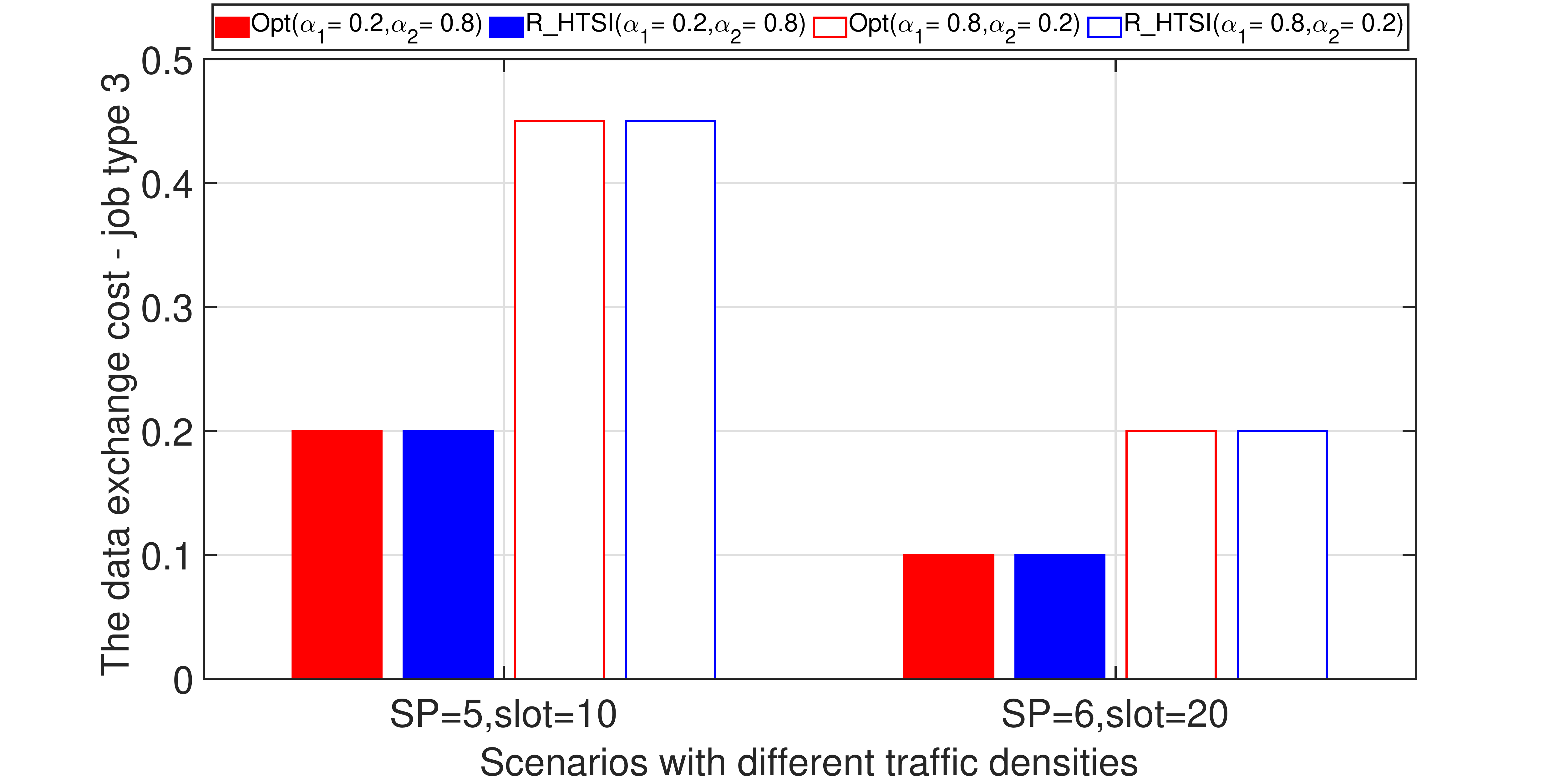}}\\
    \subfigure[]{\includegraphics[width=2.28in]{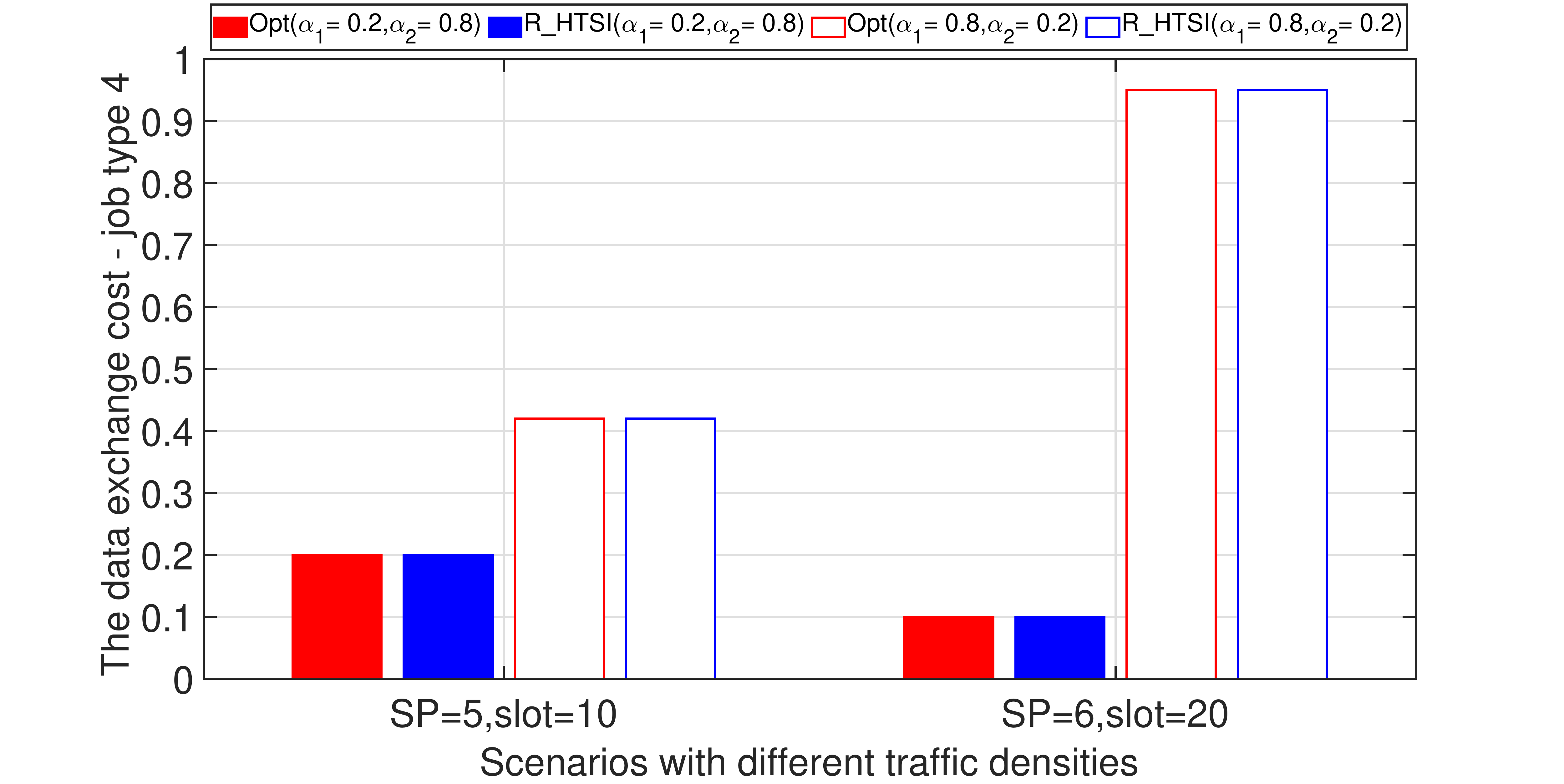}}\qquad
    \subfigure[]{\includegraphics[width=2.28in]{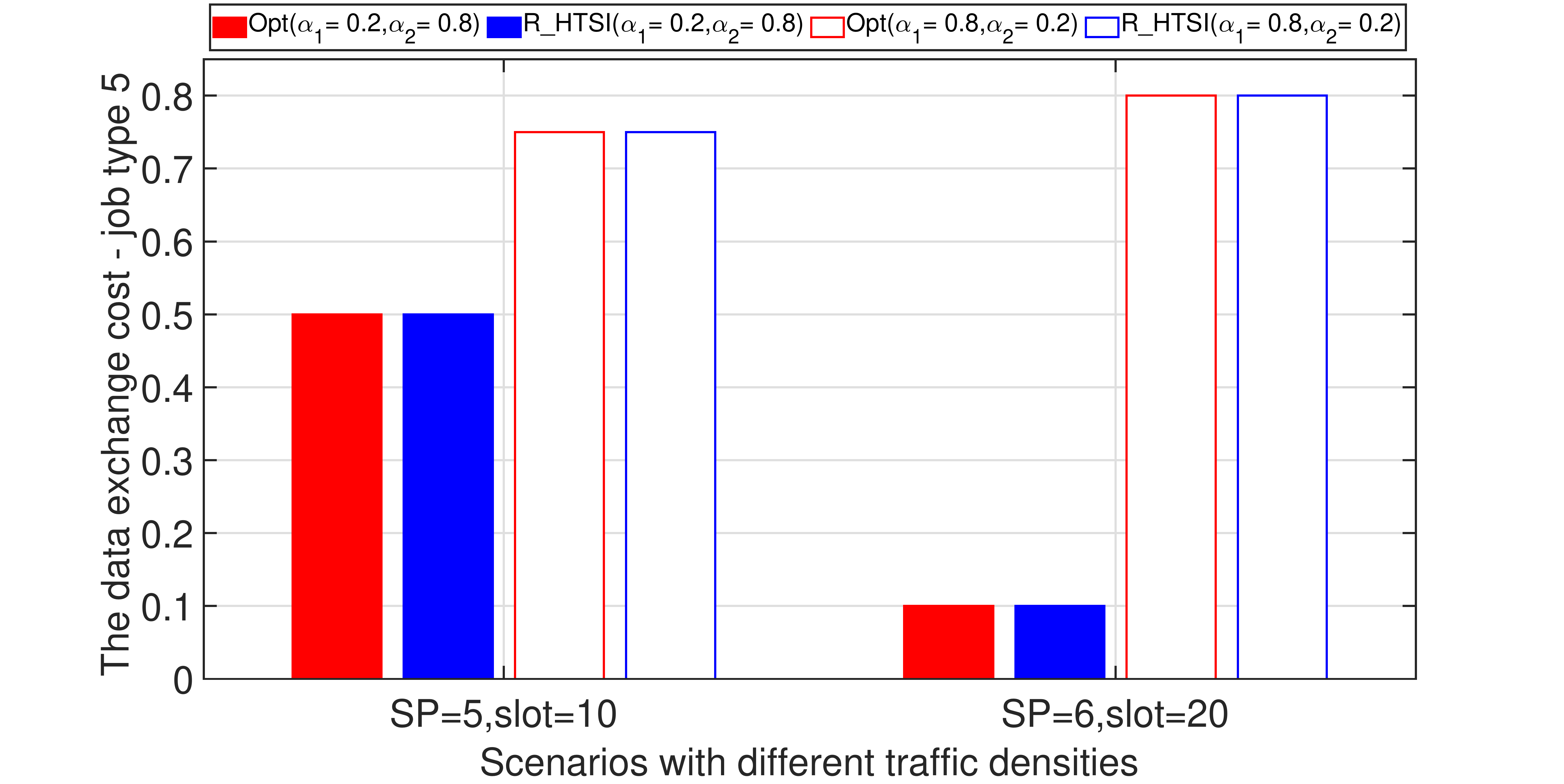}}
\caption{The data exchange cost upon selection of different values of $\alpha_1$ and $\alpha_2$ in low traffic and rush hour scenarios: a) comparison of closed triad graph jobs; b)
comparison of square graph jobs; c) comparison of bull graph jobs; d)
comparison of double-star graph jobs; e) comparison of tadpole graph jobs.}
\label{fig6}
\end{figure*}

\noindent In this section, we propose an optimal method for low-traffic scenarios, the pseudo code of which is given in \textbf{Algorithm 1}. Our method is composed of two stages: in stage 1, we find all
candidates from VC graph ${\bm G}^{{\bm S}}$ by analyzing the adjacency matrices of graph jobs and related VCs shown in step 2 and step 3; in stage 2, we select one candidate that can minimize the value of the objective function given in (4). In this method, going through
all possible candidates ensures identification of the optimal solution for the graph job
allocation; however, the computational complexity will be $O({n!}C(K,n))$, where $n$ represents the number of components in the graph job, $K=\sum _{s_j}{|{\bm \kappa_{ j}}|}$ indicates the number of available slots in the related VC, and $C(K,n)$ stands for the K-choose-n operation. As can be seen, this computational complexity is prohibitive as the vehicular density is high.

\section{Randomized Graph Job Allocation Mechanism via Hierarchical Tree based Subgraph Isomorphism}

\noindent Motivated by the intensive required computations of our previous method, we aim to develop a low-complexity graph job allocation method, to better solve larger and real-life cases.
To this end, we develop
a randomized graph job allocation algorithm by proposing a hierarchical tree based subgraph isomorphism algorithm. In our algorithm, for each graph job, we randomly choose one component and put it on an idle slot first. Then, the graph job is constructed as a hierarchical tree with the first chosen component as the root. Afterward, we randomly match each component in the hierarchical tree with one of the idle slots on the basis of different layers while satisfying all constraints shown in (4). 
Our randomized subgraph isomorphism can greatly improve the efficiency with computational-complexity of $O(n)$; moreover, the hierarchical tree based scheme ensures that components in a graph job can be matched to idle slots while satisfying the topologies of both the graph job and the vehicular cloud. The pseudo code of our algorithm is given in \textbf{Algorithm 2}. 

In this algorithm, steps 1-4 describe the initialization procedure, and steps 6--11 handle the cases in which the graph job cannot be distributed due to insufficient available slots in the related VC. In Steps 12--34, each job is seen as a hierarchical tree with layers by regarding the first randomly chosen component as the root, examples of which are depicted in Fig. 2(b). Concretely, we choose components that have one-hop connections with the root as the second layer, and components in the third layer will have one-hop connections with a component in the second layer; the remaining layers can be formed in a same manner. To satisfy edge relations between components in the hierarchical tree corresponding to a graph job, all components in the same layer should be mapped randomly into idle slots in each allocation procedure while meeting constraints (c-1), (c-2), and (c-3) shown in steps 20-30, where (c-1) ensures the data transmission between JO and SPs, (c-2) guarantees the connection between two slots on different SPs if they are assigned to handle connected components, and (c-3) ensures the contacts between SPs that deal with components in adjacent layers. Among the candidates obtained in each iteration, the best one will be reserved through comparing the value of the objective function given in (4), through steps 34-40.

\section{Numerical Results and Performance Evaluation}

\noindent This section presents numerical results illustrating the performance of the
proposed mechanisms. For the convenience of analysis, the optimal solution and the randomized solution via hierarchical tree based subgraph isomorphism are abbreviated as ``Opt`` and ``R$\_$HTSI``, respectively. The graph job topologies considered in simulations are depicted in Fig. 2(a). We assume
$\varepsilon =\xi =0.9$, $t_{jm}^{Trans}\mathrm{\in [0.2,0.6]}$,
$c_{jj'}^{Exch}\mathrm{\in [0.2,0.6]}$, $\omega_{ii'}^{A}\mathrm{\in
[0.1,0.4]}$, and $\lambda_{jj'}\in [0.01,0.06]$, where each parameter is randomly chosen in its respective interval.

Considering the graph job types shown in Fig. 2(a), Fig. 3 depicts the running time performance comparison between the optimal
and the proposed R$\_$HTSI mechanisms for different sizes and configurations of the VC network. Since the gap between the running time of the R$\_$HTSI and the optimal mechanism becomes too large as the graph job structure becomes more complicated (i.e., contains more nodes and edges), 10-based logarithm of the running time is presented in the left sub-plot of Fig. 3(b), Fig. 3(c), Fig. 3(d) and Fig. 3(e). Compared with the 
R$\_$HTSI, the left sub-plot in each sub-figure indicates that as the traffic density and the number of available slots of SPs in a VC grow, the running time of the optimal mechanism may rise sharply, which makes it unsuitable for fast-changing and large IoV networks. Notably, different topological complexity of VC configurations (e.g., existence of more service providers and connections between vehicles in VC graph) may lead to a dramatic change in the running time performance. For example, the left sub-plot of Fig. 3(c) shows that the running time of the optimal mechanism increases approximately 10 times when the number of SPs increases from 4 to 5 with the average number of slots per SP equal to 4. Also, in the left sub-plot of Fig. 3(d), considering the number of SPs to be 5, as the average number of slots per SP grows from 3 to 5, the running time of the optimal mechanism experiences about a hundredfold increase, while that of the proposed R$\_$HTSI remains almost unchanged. Note that a small gap in the running time of the optimal mechanism upon increasing the size of the network in some scenarios, e.g., in the left sub-plot of Fig. 3(a) upon increasing the number of SPs from 7 to 8 when the average number of slots per SP equals to 6, is due to the possibility of the existene of more edges between the vehicles in smaller VCs. 
The right sub-plot of each sub-figure depicts the running time of the R$\_$HTSI considering different number of iterations. Comparing the values depicted in the right sub-plot with the running time of the optimal mechanism in the left subplot, using the proposed R$\_$HTSI mechanism results in significant decrease in the running time of graph job allocation. For instance, the utilizing of the R$\_$HTSI in Fig. 3(d) can achieve about 4500 times running time reduction when the number of SPs and slots per SP are both equal to 5, and offers approximately 3 thousand fold decrease in the running time on average, as compared to the optimal mechanism. Moreover, Fig. 3(e) also shows about 3 thousand times running time reduction on average upon utilizing the R$\_$HTSI mechanism. The right sub-plot in each sub-figure reveals that the running time of the proposed R$\_$HTSI mechanism remains low as the number of
iterations increases, allowing this mechanism to be implemented efficiently
in rapidly changing and large-scale IoV networks, particularly during rush hours.

By utilizing Monte-Carlo iterations, the comparisons of the objective
function values (given in (4)) between the optimal and the R$\_$HTSI mechanisms for
different graph job types and vehicular cloud sizes as well as configurations are shown in Fig. 4 with
$\alpha_1=\alpha_2=0.5$. For each type of graph job, low-traffic
and rush-hour cases reveal that as the number of iterations increases, the performance of the R$\_$HTSI mechanism
can approach the optimal one with a much lower computational complexity,
which is more applicable in large and real-life IoV scenarios.

Fig. 5 and Fig. 6 depict the job completion time and data exchange cost upon using different values of $\alpha_1$ and $\alpha_2$, respectively, under low-traffic and rush hour scenarios. As can be seen from Fig. 5, for each graph job type, the larger the value of $\alpha_1$ is, the more sensitive the JO will be to minimize the job completion time rather than the data exchange cost. On the contrary, larger $\alpha_2$ leads to a lower data exchange cost as illustrated in Fig. 6 under various traffic scenarios.

\section{Conclusion, Challenges, and Future Work}

\noindent This paper studies novel allocation mechanisms for computation-intensive
graph jobs over VCs via opportunistic V2V communications in IoV frameworks, which is modeled as a nonlinear integer programming problem with constraints.
For low-traffic scenarios, an optimal approach is introduced. For rush-hour scenarios, we propose a randomized graph job allocation mechanism via hierarchical tree based subgraph isomorphism with
low complexity, which efficiently addresses the graph job allocation problem in
large IoV networks. Based on comprehensive simulations, the effectiveness of the proposed methods in
low-traffic scenarios is revealed. In high-traffic
scenarios, we demonstrate a significant running time gap between the optimal
and proposed sub-optimal R$\_$HTSI mechanisms; moreover, sub-optimal solutions can
closely approach the optimal ones as the number of iterations increases.

Several challenges remain in the context of graph job allocation in IoV
environments, such as considering load
balancing issues so as to improve on-board resource utilizations, and designing more efficient subgraph isomorphism algorithms to accommodate fast changing topologies in the IoV environment. 
Furthermore, increasingly complex traffic conditions and
job types lead to growing challenges for low-complexity mechanisms in
larger networks. One main goal of our future work is to explore more efficient models for designing graph job allocation
mechanisms in IoV, while ensuring load balancing and low computation complexity.

\section*{Acknowledgment}

\noindent This work is supported in part by the 2015 National Natural Science
Foundation of China (grant no. 61401381), the Major Research Plan of the
National Natural Science Foundation of China (grant no. 91638204), the State
Key Program of the National Natural Science Foundation of China (grant no.
61731012), 2018 National Natural Science Foundation of China (grant no.
61871339), Digital Fujian Province Key Laboratory of IoT Communication,
Architecture and Safety Technology (grant no. 2010499) and the US National
Science Foundation under Grants ECCS-1444009 and CNS-1824518.

\ifCLASSOPTIONcaptionsoff
  \newpage
\fi



%\bibliographystyle{IEEEtran}

\begin{thebibliography}{24}


\bibitem{1} N. Zhang et al., ``Software defined space-air-ground integrated vehicular
networks: challenges and solutions,'' \textit{IEEE Commun. Mag.}, vol. 55, no.7, 2017, pp. 101--109.

\bibitem{2} D. Zhang et al., ``A Power Allocation-Based Overlapping Transmission Scheme in Internet of Vehicles,'' \textit{IEEE Internet Things J.}, vol. 6, no. 1, 2019, pp. 50--59.

\bibitem{3} M. Chao et al., ``Data-Driven State-Increment Statistical Model and Its
Application in Autonomous Driving,'' \textit{IEEE Trans. Intell. Transp. Syst.}, 2018.

\bibitem{4}C. Yu et al., ``Deployment and Dimensioning of Fog Computing-based Internet of Vehicle Infrastructure for Autonomous Driving,'' \textit{IEEE Internet Things J.}, 2018, pp. 1--1.

\bibitem{5} I. Parvez et al., ``A survey on low latency towards 5G: RAN, core
network and caching solutions,'' \textit{IEEE Commun. Surveys Tut.},
2018.

\bibitem{6} D. Huang et al., ``A dynamic offloading algorithm for mobile
computing,'' \textit{IEEE Trans. Wireless Commun.}, vol.11, no. 6,
2012, pp. 1991--1995.

\bibitem{7} M. Jia et al., ``Heuristic offloading of concurrent tasks for computation-intensive applications in mobile cloud computing,'' \textit{ Int’l. Conf. 2014 IEEE INFOCOM WKSHPS}, Toronto, CA, Apr. 2014, pp. 352--357.

\bibitem{8} S. Hosseinalipour et al., ``Allocation of
graph jobs in geo-distributed cloud networks,'' arXiv preprint
arXiv:1808.04479, 2018.

\bibitem{9} J. Ghaderi et al., ``Scheduling storms and streams in the cloud,'' \textit{ACM
Trans. Modeling and Performance Eval. of Comput. Syst.}
, vol. 1, no. 4, 2016, pp. 1--14.

\bibitem{10} Y. Mao et al., ``Joint task offloading scheduling and transmit power allocation for mobile-edge computing systems,'' \textit{ Int’l. Conf. 2017 IEEE WCNC}, San Francisco, USA, Mar. 2017, pp. 1--6.

\bibitem{11} Z .Ning et al., ``A cooperative partial computation offloading scheme for mobile edge computing enabled Internet of Thing,'' \textit{IEEE Internet Things J.}, 2018, pp. 1--1.

\bibitem{12} T. Taleb et al., ``On multi-access edge computing: A survey of the
emerging 5G network edge cloud architecture and orchestration,'' \textit{IEEE Commun. Surveys Tut.}, vol. 19, no. 3, 2017, pp. 1657--1681.

\bibitem{13} T. Mekki et al., ``Vehicular cloud networks: Challenges, architectures,
and future directions,'' \textit{Veh. Commun.}, vol. 9, 2017, pp. 268--280.

\bibitem{14} D. Xu et al., ``A survey of opportunistic offloading,'' \textit{IEEE Commun. Surveys Tut.}, 2018.

\bibitem{15} G.-P. A. W. Group, ``A study on 5g v2x deployment,''
https://5g-ppp.eu/wp
content/uploads/2018/02/5G-PPP-Automotive-WG-White-P-\linebreak\\[-8pt]aper Feb.2018.pdf, 2018.

\bibitem{16} Y. Wu et al., ``Secrecy-Driven Resource Management for Vehicular
Computation Offloading Networks,'' \textit{IEEE Netw.},~vol. 32, no. 3, 2018,
pp. 84--91.

\bibitem{17} J. Zhou et al., ``Reliability-oriented optimization of computation offloading for cooperative vehicle-infrastructure systems,'' \textit{ IEEE Signal Process. Lett.}, vol. 26, no. 1, 2019, pp. 104--108.

\bibitem{18} K. Zhang et al., ``Optimal delay constrained offloading for vehicular edge computing networks,'' \textit{ Int’l. Conf. 2017 IEEE ICC}, Paris, France, May. 2017, pp. 1--6.

\bibitem{19} Y. Liu et al., ``A Computation Offloading Algorithm Based on Game Theory for Vehicular Edge Networks,'' \textit{ Int’l. Conf. 2018 IEEE ICC}, Kansas City, usa, May. 2018, pp. 1--6.

\bibitem{20} L. Shi et al., ``Energy-aware scheduling of embarrassingly parallel
jobs and resource allocation in cloud,'' \textit{IEEE Trans. Parallel Distrib. Syst.}, vol. 28, no. 6, 2017, pp. 1607--1620.

\bibitem{21} Goudarzi. M et al., ``A fast hybrid multi-site computation offloading for mobile cloud computing,'' \textit{Journal of Netw. Comput. Appl.}, vol. 66, 2017, pp. 219--231.

\bibitem{22} Cordella .L P et al., ``A (sub) graph isomorphism algorithm for matching large graphs,'' \textit{IEEE Trans. Pattern Anal. Mach. Intell.}, vol. 26, no. 10, 2004, pp. 1367--1372.

\bibitem{23} X. Zhu et al., ``Contact-aware optimal resource allocation for mobile
data offloading in opportunistic vehicular networks,'' \textit{IEEE Trans. Veh. Technol.}, vol. 66, no. 8, 2017, pp. 7384--7399.

\bibitem{24} H. Guo et al., ``Mobile edge computation offloading for ultradense IoT networks,'' \textit{IEEE Internet Things J.}, vol. 5, no. 6, 2018, pp. 4977--4988.

\end{thebibliography}

\vspace*{-0.7cm}

%\begin{IEEEbiographynophoto}{Minghui LiWang}
%(minghuilw@stu.xmu.edu.cn) received her B.S. degree in
%Computer Science and Technology in Huaqiao University, Xiamen, China, 2013 and is on a master-doctor continuous
%study course, currently a Ph.D candidate in the School of Information Science and Engineering, Xiamen
%University, China. She was a visiting Ph.D at NC State University during 2017-2018. Her
%research interests are wireless communication systems, cloud computing, and
%Internet of Vehicles.
%\vspace*{-0.8cm}

\begin{IEEEbiography}[{\includegraphics[width=1in,height=1.in,clip,keepaspectratio]{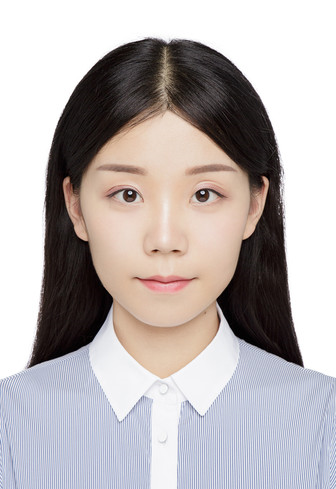}}]{Minghui LiWang}
	(minghuilw@stu.xmu.edu.cn) received her B.S. degree on 
	computer science and technology in 2013 and started a 
	master-doctor continuous study course in 2015, currently a Ph.D candidate in the 
	Department of Communication Engineering, School of Information Science and 
	Engineering, Xiamen University, China. She is now a visiting Ph.D at NC 
	State University. Her research interests are wireless communication systems, 
	cloud computing, and Internet of Vehicles.
\end{IEEEbiography}
\vspace*{-3em}

\begin{IEEEbiography}[{\includegraphics[width=1in,height=1.in,clip,keepaspectratio]{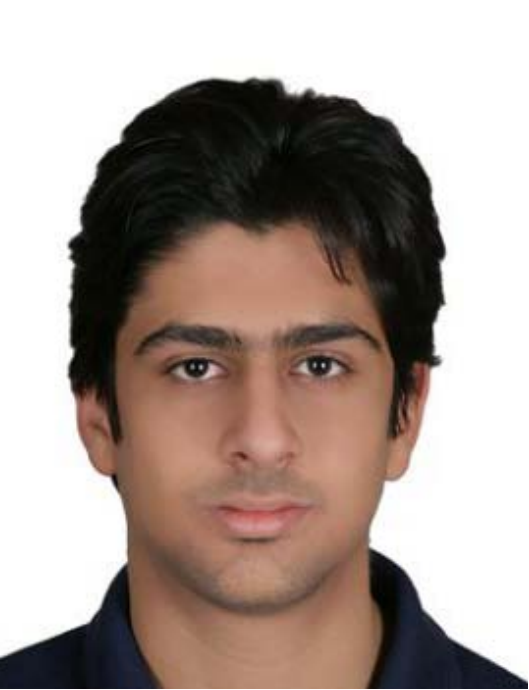}}]{Seyyedali Hosseinalipour}
(shossei3@ncsu.edu) received his B.S. degree in
Electrical Engineering from Amirkabir University of Technology (Tehran
Polytechnic), Tehran, Iran in 2015. He is pursuing a Ph.D. degree in the
Department of Electrical and Computer Engineering at North Carolina State
University, Raleigh, NC, USA. His research interests include analysis of
wireless networks, resource allocation and load balancing for cloud
networks, and resource allocation and task scheduling for vehicular ad-hoc
networks.
\end{IEEEbiography}
%\vspace*{-1cm}

\begin{IEEEbiography}[{\includegraphics[width=1in,height=1.in,clip,keepaspectratio]{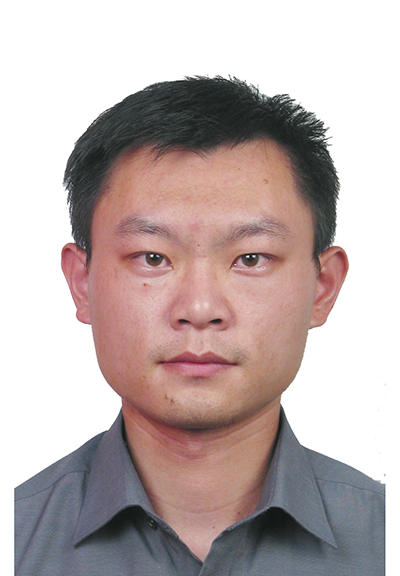}}]{Zhibin Gao}
(gaozhibin@xmu.edu.cn) received his B.S. degree in Communication
Engineering in 2003, M.S. degree in Radio Physics in 2006, and Ph.D. in
Communication Engineering in 2011 from Xiamen University, where he is a
senior engineer of communication engineering. His current research interests
include wireless communication, mobile network resource management, and
signal processing.
\end{IEEEbiography}
%\vspace*{-1cm}

\begin{IEEEbiography}[{\includegraphics[width=1in,height=1.in,clip,keepaspectratio]{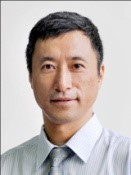}}]{Yuliang Tang}
(tyl@xmu.edu.cn) received his M.S. degree from Beijing
University of Posts and Telecommunications, China in 1996. He received his
Ph.D. degree in Communication Engineering from Xiamen University in 2009,
where he is a professor in the Department of Communication Engineering. His
research interests include wireless communication, 5G, and vehicular ad-hoc
networks.
\end{IEEEbiography}
%\vspace*{-1cm}

\begin{IEEEbiography}[{\includegraphics[width=1in,height=1.in,clip,keepaspectratio]{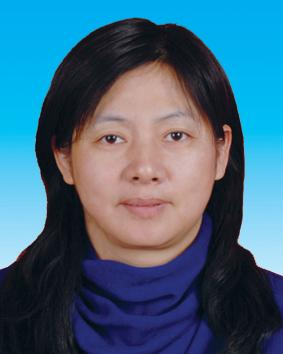}}]{Lianfen Huang}
(lfhuang@xmu.edu.cn) received her B.S. degree in Radio
Physics in 1984 and PhD in Communcation Engineering in 2008 from Xiamen
University. She was a visiting scholar in Tsinghua University in 1997. She
is a professor in Department of Communication Engineering, Xiamen
University, Xiamen, Fujian, China. Her current research interests include
wireless communication, wireless network and signal process.
\end{IEEEbiography}
%\vspace*{-1cm}

\begin{IEEEbiography}[{\includegraphics[width=1in,height=1.in,clip,keepaspectratio]{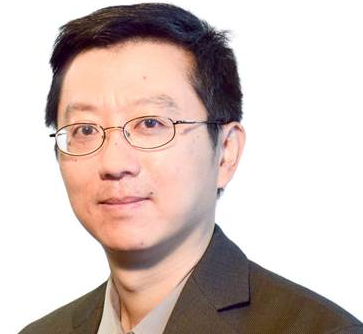}}]{Huaiyu Dai}
[F'17] (hdai@ncsu.edu) received the B.E. and M.S. degrees in
electrical engineering from Tsinghua University, Beijing, China, in 1996 and
1998, respectively, and the Ph.D. degree in electrical engineering from
Princeton University, Princeton, NJ in 2002. He was with Bell Labs, Lucent
Technologies, Holmdel, NJ, in summer 2000, and with AT{\&}T Labs-Research,
Middletown, NJ, in summer 2001. He is currently a Professor of Electrical
and Computer Engineering with NC State University, Raleigh. His research
interests are in the general areas of communication systems and networks,
advanced signal processing for digital communications, and communication
theory and information theory. His current research focuses on networked
information processing and crosslayer design in wireless networks, cognitive
radio networks, network security, and associated information-theoretic and
computation-theoretic analysis. He has served as an editor of IEEE
Transactions on Communications, IEEE Transactions on Signal Processing, and
IEEE Transactions on Wireless Communications. Currently he is an Area Editor
in charge of wireless communications for IEEE Transactions on
Communications. He co-edited two special issues of EURASIP journals on
distributed signal processing techniques for wireless sensor networks, and
on multiuser information theory and related applications, respectively. He
co-chaired the Signal Processing for Communications Symposium of IEEE
Globecom 2013, the Communications Theory Symposium of IEEE ICC 2014, and the
Wireless Communications Symposium of IEEE Globecom 2014. He was a
co-recipient of best paper awards at 2010 IEEE International Conference on
Mobile Ad-hoc and Sensor Systems (MASS 2010), 2016 IEEE INFOCOM BIGSECURITY
Workshop, and 2017 IEEE International Conference on Communications (ICC
2017).
\end{IEEEbiography}

\vfill

\end{document}